\documentclass[superscriptaddress,twocolumn,floatfix]{revtex4-1}
\usepackage{amssymb,amsmath,amsfonts,amsthm,dsfont}
\usepackage[normalem]{ulem}
\usepackage{bbm}
\usepackage{physics}
\usepackage{graphicx,wasysym}
\usepackage{xcolor}
\usepackage{float}
\usepackage{mathtools}
\usepackage[ruled,vlined]{algorithm2e}

\usepackage[caption=false]{subfig}

\DeclarePairedDelimiter{\ceil}{\lceil}{\rceil}

\newtheorem{theorem}{Theorem}
\newtheorem{proposition}{Proposition}

\renewcommand{\Vec}[1]{\mathbf{#1}}
\newcommand{\iu}{{i\mkern1mu}}

\newcommand{\RR}{\mathbbm{R}}
\newcommand{\NN}{\mathbbm{N}}

\newcommand{\CC}{\mathbbm{C}}

\usepackage{xifthen}% provides \isempty test
\newcommand{\PP}[1][]{
  \ifthenelse{\isempty{#1}}
    {\mathbbm{P}}
    {\mathbbm{P}\left[#1\right]}
}
\newcommand{\EE}[1][]{
  \ifthenelse{\isempty{#1}}
    {\mathbbm{E}}
    {\mathbbm{E}\left[#1\right]}
}

\newcommand{\sumt}[1]{\mathrm{sum}\left\{#1\right\}}

\begin{document}
\title{Quantum-enhanced analysis of discrete stochastic processes}

\author{Carsten Blank}
\email{blank@data-cybernetics.com }
\affiliation{Data Cybernetics, 86899 Landsberg, Germany}
\author{Daniel K. Park}
\email{dkp.quantum@gmail.com}
\affiliation{School of Electrical Engineering, KAIST, Daejeon, 34141, Republic of Korea}
\affiliation{ITRC of Quantum Computing for AI, KAIST, Daejeon, 34141, Republic of Korea}
\author{Francesco Petruccione}
\email{petruccione@ukzn.ac.za}
\affiliation{School of Electrical Engineering, KAIST, Daejeon, 34141, Republic of Korea}
\affiliation{Quantum Research Group, School of Chemistry and Physics, University of KwaZulu-Natal, Durban, KwaZulu-Natal, 4001, South Africa}
\affiliation{National Institute for Theoretical Physics (NITheP), KwaZulu-Natal, 4001, South Africa}

% \begin{abstract}
% Discrete stochastic processes are instrumental for modelling the dynamics of probabilistic systems and have a wide spectrum of applications in science and engineering. Simulations by Monte Carlo methods are powerful for analysis while keeping the exponential size of the full process at bay. Two fundamental theorems give the basis of this method: the law-of-large-numbers and the central-limit-theorem. However, high variances of the estimates are the drawback to the method. Importance sampling was developed over the years to overcome these difficulties. Here we show that quantum computers can enhance the analysis of such processes by reducing the Monte Carlo sampling to a standard Bernoulli trial -- a coin toss essentially. This has the favorable effect of setting the sampling variance to an optimum. We exploit recent discoveries reporting a quadratic speed-up using a method lent from the well known Grover search -- namely amplitude estimation -- and show that both methods allow for reduction of sampling variance and a speed-up of the convergence. These new developments make a stronger case for quantum computers in applications. We investigate the feasibility of the method with proof-of-principle experiments that are performed using the IBM quantum cloud platform.
% \end{abstract}

\begin{abstract}
Discrete stochastic processes (DSP) are instrumental for modelling the dynamics of probabilistic systems and have a wide spectrum of applications in science and engineering. DSPs are usually analyzed via Monte Carlo methods since the number of realizations increases exponentially with the number of time steps, and importance sampling is often required to reduce the variance. We propose a quantum algorithm for calculating the characteristic function of a DSP, which completely defines its probability distribution, using the number of quantum circuit elements that grows only linearly with the number of time steps. The quantum algorithm takes all stochastic trajectories into account and hence eliminates the need of importance sampling. The algorithm can be further furnished with the quantum amplitude estimation algorithm to provide quadratic speed-up in sampling. Both of these strategies improve variance beyond classical capabilities. The quantum method can be combined with Fourier approximation to estimate an expectation value of any integrable function of the random variable. Applications in finance and correlated random walks are presented to exemplify the usefulness of our results. Proof-of-principle experiments are performed using the IBM quantum cloud platform.
\end{abstract}

\maketitle
\def\one{{\mathchoice {\rm 1\mskip-4mu l} {\rm 1\mskip-4mu l} {\rm \mskip-4.5mu l} {\rm 1\mskip-5mu l}}}
% I've re-inserted the section numbers for the arXiv version (and we can also submit to a journal as it is).
% ====================================================================
\section{Introduction}
% ====================================================================

Simulation of physical processes on quantum computers~\cite{Feynman:1981tf,10.2307/2899535} has many facets, with recent developments improving the usage of this technology~\cite{doi:10.1146/annurev-physchem-032210-103512,zalka1998simulating,blatt2012quantum,barreiro2011open,gerritsma2010quantum,aspuru2012photonic,friedenauer2008simulating,weimer2010rydberg,aspuru2005simulated,lanyon2010towards,abrams1999quantum,berry2007efficient}. While one obvious task for a quantum computer is to simulate quantum mechanical behaviors of nature~\cite{Feynman:1981tf}, quantum simulation of probability distributions and stochastic processes has gained attention recently~\cite{rebentrost2018quantumMC, rebentrost2018quantumMC, woerner2019quantum,heinrich2003monte,heinrich2003monte,gu2012quantum,ghafari2019dimensional,ghafari2019interfering}. Since quantum mechanics can be viewed as a mathematical generalization of probability theory, where non-negative real-valued probabilities are replaced by complex-valued probability amplitudes, quantum computing appears to be a natural tool for simulating classical probabilistic processes. Intuitively, some quantum advantage is expected since the probability amplitudes can interfere, unlike in the classical probabilistic computing. Indeed, a quadratic quantum speed-up has been reported for solving financial problems when compared to Monte Carlo simulations~\cite{woerner2019quantum}.

Although the main focus of this work is to achieve quantum sampling advantage for discrete stochastic processes (DSPs), we note in passing that quantum memory advantage for stochastic processes with causal structures has been shown elsewhere~\cite{gu2012quantum,ghafari2019dimensional,ghafari2019interfering}.

Classically, Monte Carlo methods are essential for estimating expected values of random variables in DSPs, since the number of realizations increases exponentially with the number of time steps. When the Monte Carlo sampling is repeated $N$ times, the expectation value to be found converges with $\mathcal{O}(1/\sqrt{N})$ regardless of the number of realizations. The central limit theorem ensures this, but it is important to note that this convergence is only attained in the limit of $N \rightarrow \infty$. When this limit is not nearly attained, it is often crucial to have sampling strategies to reduce the variance of the estimate. The reason for this is that less likely events are also less likely to be sampled from, while such events can be of high impact. This can create a bias in the computation as the estimation can be dominated by more probable, but less important values. To correctly sample with a given $N$ that is not near the limit of the large number, it is beneficial to modify the probability of the process in a way that balances the importance of the events. This is called \textit{importance sampling}~\cite{blanchet2008state,hastings1970monte,rubinstein2016simulation,Owen98safeand,cappe2008adaptive,srinivasan2013importance,martino2017layered}. This makes apparent two problems with Monte Carlo algorithms. First, it may be difficult to calculate the probability of a particular realization of a random variable. Second, importance sampling strategy relies on sophisticated understanding of the probability distribution. 

In this manuscript we show that the characteristic function of a DSPs can be efficiently calculated on a quantum computer, and in so doing we introduce an effect that can be called \textit{quantum brute-force}. The random variables of the DSP of interest do not need to be identically and independently distributed, and non-Markovian processes can also be studied. The quantum state space that grows exponentially with the number of qubits is used to move along all paths of the discrete stochastic process simultaneously in quantum superposition, and hence the term quantum brute-force is adequate. 

With a Pauli measurement scheme on a single qubit, essentially the probability of a Bernoulli trial needs to be estimated which exhibits the optimal variance that can be achieved according to the central limit theorem. This leads to a crucial result that no sampling strategies are necessary. Moreover, we connect recent developments in quantum finance~\cite{rebentrost2018quantumMC, rebentrost2018quantumPortfolio, woerner2019quantum} to our method and show that sampling convergence can be improved by means of quantum amplitude estimation (AE) by a power of two. The methods introduced here point therefore to an exciting and promising use of quantum computers: less variance and faster convergence for Monte-Carlo sampling.

Applications of our method span extensively across many fields in science and engineering; any physical behaviour that can be mathematically modelled as a DSP can be studied in principle. In particular, we show how the found simulation procedure can be applied to option pricing theory and to correlated random walks which leads to various applications in biology, ecology and finance. For each example, we experimentally demonstrate the proof-of-principle using the IBM quantum cloud platform.

% ====================================================================
\section{Results}
% ====================================================================
A discrete stochastic process can be described with $n$ discrete random variables $X_l: \Omega_l \rightarrow \RR$, $l = 1, \ldots, n$ for some $n \in \NN_+$, each having at most $k$ non-zero realizations, i.e. given a sample space $\Omega_l$, there exist at most $k$ elements $x_{l,0}, \ldots, x_{l,k-1} \in \RR$ with $X_l(\Omega_l) = \{x_{l,0}, \ldots, x_{l,k-1}\}$. Hereinafter, we use the following notations. Each realization of the stochastic process is identified with an index vector $\Vec{j} = (j_1, \ldots, j_n)^\top \in K^n$ with $K := \{0, \ldots, k - 1\}$ and is denoted by $\Vec{x}(\Vec{j}) = (x_{1,j_1}, \ldots, x_{n,j_n})^\top$. Moreover, $\sumt{\Vec{x}(\Vec{j})} = \sum_{l=1}^n x_{l,j_l}$ and $\Vec{x}^{(m)}(\Vec{j}) = (x_{1,j_1}, \ldots, x_{m,j_m})^\top$ with $m\leq n$. The first and the second subscripts of the random variables and probabilities label the time step and the event, respectively. A quantity of interest for such processes is the expectation value of an integrable function $f: \RR \rightarrow \RR$ of the random variable $S_n = \sum_{l=1}^n X_l$
\begin{equation}
\label{eq:expectation_value_joint_prob}
\EE[f(S_n)] = \sum_{\Vec{j} \in K^n} f(\sumt{\Vec{x}(\Vec{j})}) \PP[\Vec{X} = \Vec{x}(\Vec{j})].
\end{equation}

Now, we explain how to encode the described stochastic process in a quantum state, and evaluate equation~(\ref{eq:expectation_value_joint_prob}) by making a measurement on the quantum state. The quantum state consists of an index system and a data system defined in a Hilbert space $\mathcal{H}_\mathcal{I}\otimes\mathcal{H}_\mathcal{D}=\mathbb{C}^{k^n}\otimes\mathbb{C}^{d}$, where $\mathcal{I}$ ($\mathcal{D}$) indicates the index (data) system and $d$ is determined by the problem of interest. Each realization of the stochastic process is represented as a unitary operator $U(\cdot): \RR^n \longrightarrow B(\mathcal{H}_\mathcal{I} \otimes \mathcal{H}_\mathcal{D})$ parametrized by some $n$-dimensional vector ($B(\mathcal{H})$ is the space of linear operators on a Hilbert space $\mathcal{H})$. Then a DSP can be represented as
\begin{equation}
\label{eq:process_state}
    \ket{\Psi_f} = \sum_{\Vec{j} \in K^n} p(\Vec{j}) \ket{\Vec{j}} \otimes U(\Vec{j}) \ket{\psi}.
\end{equation}
The factor of each part of the sum are denoted by $p(\Vec{j})$ with $\sum_{\Vec{j}} p^2(\Vec{j}) = 1$ and the state of the index system is denoted by $\ket{\Vec{j}} = \ket{j_n \cdots j_1} = \ket{j_n} \otimes\cdots\otimes \ket{j_1}$, where $\ket{j_l}$ for $j_l = 0, \ldots, k - 1$ is the orthogonal basis. As described in Ref.~\cite{Park_2019_forking_sampling}, measuring an expectation value of an observable $M$ on the data system of the final state in equation~(\ref{eq:process_state}) yields the convex sum of independent expectation values measured from all $k^n$ trajectories as
\begin{align}
    \expval{M} &= \bra{\Psi_f} I_{\mathcal{I}} \otimes M \ket{\Psi_f} \nonumber \\
    &= \sum_{\Vec{j} \in K^n} p^2(\Vec{j}) \bra{\psi} U^\dagger(\Vec{j}) M U(\Vec{j}) \ket{\psi} \nonumber \\
    &= \sum_{\Vec{j} \in K^n} p^2(\Vec{j}) \expval{M(\Vec{j})}_\psi, \label{eq:forking_expval}
\end{align}
where $M(\Vec{j}) = U^\dagger(\Vec{j}) M U(\Vec{j})$. The coefficient $p^2(\Vec{j})$ can be identified with the joint probability and the expectation value with the evaluation of $f$, i.e.
\begin{align}
    p^2(\Vec{j}) &\equiv \PP\left[X_1 = x_{1,j_1}, \ldots, X_n = x_{n,j_n}\right] \label{eq:join_probability_identification} \\
    \expval{M(\Vec{j})} &\equiv f(x_{1,j_1} + \cdots + x_{n,j_n}) \label{eq:function_evaluation_identification}
\end{align}
for a function $f: \RR \rightarrow \RR$ that we will specify below.

In the worst case, evaluation of equation~(\ref{eq:forking_expval}) requires two expensive procedures as follows. First, $k^n$ probabilities need to be encoded as the amplitudes of $k^n$ computational basis given by $n$ qudits of dimension $k$. This can be done with various quantum state preparation techniques with substantial amount of computational overhead~\cite{Mottonen:2005:TQS:2011670.2011675,ffqram}. Next, $k^n$ unitary operators, conditioned on all possible index states, need to be applied on an input state $|\psi\rangle$. Such operators can be expressed as
\begin{align*}
    c\text{-}U(\Vec{j}) = &\ketbra{\Vec{j}}{\Vec{j}} \otimes \left(V(x_{n,j_n}) \cdots V(x_{1,j_1})\right) + \ketbra{\Vec{j}}{\Vec{j}}_\perp \otimes I_{\mathcal{D}}
\end{align*}
with $V: \RR \rightarrow \mathcal{H}_{\mathcal{D}}$. On the other hand, for many interesting DSPs, the number of necessary unitary operators can be reduced to $\mathcal{O}(nk)$. 

Before explaining such exponential-reduction in detail, we introduce two results in the next two propositions (with proofs provided in Supplementary Information) to establish the grounds for the measurement scheme.
\begin{proposition}[Pauli X and Y Measurement]\label{prop:1}
Let $V(x) = R_z(2x) = \ketbra{0}{0} + e^{\iu x} \ketbra{1}{1}$ with $x\in\RR$, then by setting $M=\sigma_x$ or $M=\sigma_y$ and $\ket{\psi} = (\ket{0} + \ket{1})/\sqrt{2}$, we find
\begin{align}
    \expval{I_{\mathcal{I}} \otimes \sigma_x}_{\Psi_f} &= \EE[\cos(S_n)] \label{eq:expval_sigma_x}, 
    \\
    \expval{I_{\mathcal{I}} \otimes \sigma_y}_{\Psi_f} &= \EE[\sin(S_n)]. \label{eq:expval_sigma_y}
\end{align}
\end{proposition}
As the true value of $\EE[\cos(S_n)]$ and $\EE[\sin(S_n)]$ must be estimated, in general the convergence behaves according to the central limit theorem, which guarantees that the measurement statistics approaches to a normal distribution around a mean that corresponds to $\EE[f(S_n)]$ as the number of experiments $N_S \rightarrow \infty$. The speed of convergence is moreover given by $\mathcal{O}(1/\sqrt{N_S})$. Taking into account that the Pauli measurements have two eigenvalues, the task is essentially estimating a probability of a Bernoulli trial, which specifies how the central limit theorem is realized. Given a confidence of $1-\alpha$, the number of experiments to be within a margin of error $\epsilon>0$ is $N_S = \ceil{z_\alpha^2/(4\epsilon^2)}$ with $z_\alpha = z(1 - \alpha/2)$ being the quantile function. In contrast, with classical Monte Carlo sampling the convergence rate by the central limit theorem is achieved with the caveat that the Monte Carlo simulation samples from an usually unknown stochastic process and concise estimates on the number of experiments given a margin of error are in general not easily accessible. As we see, the property that equation~(\ref{eq:process_state}) encompasses \textit{all} possible paths with the correct probability with a quantum measurement leads to the seemingly small advantage of knowing the convergence before hand, irrespective of the distribution of the underlying DSP. Contemplation on this fact reveals that this is no small feat: the quantum advantage lies in the fact that no sampling strategies are necessary.

The quantum amplitude estimation algorithm~\cite{brassard2002quantum} can provide further speedup compared to the convergence rate of classical Monte Carlo method given by the central limit theorem, as suggested in Refs.~\cite{doi:10.1098/rspa.2015.0301,woerner2019quantum,rebentrost2018quantumMC,rebentrost2018quantumPortfolio}. The algorithm uses $m$ ancilla qubits in addition to data and index qubits and $\mathcal{O}(\mathrm{poly}(m))$ number of Grover-like iterators followed by $\mathcal{O}(m^2)$ Hadamard and controlled phase-shift gates for quantum fourier transform (QFT)~\cite{Nielsen:2011:QCQ:1972505} to estimate an amplitude with an error of $\mathcal{O}(1/2^m)$. The number of repeated measurements needed to reach confidence that the estimation succeeded is independent of $m$. The quantum AE algorithm can be adapted to our method to construct an even more powerful strategy by formulating an AE problem as follows. Given the final state in equation~(\ref{eq:process_state}) written as a linear combination $\ket{\Psi_f} = \ket{\Psi_0} + \ket{\Psi_1}$ by separating the full Hilbert space into two orthogonal subspaces, we estimate the amplitude defined as $a = \braket{\Psi_1}{\Psi_1}$. Then, the following proposition connects AE with the DSP simulation.
\begin{proposition}[Amplitude Estimation]\label{prop:2}
Let $V(x) = R_y(x) = \cos(x/2) I - \iu \sin(x/2) \sigma_y$, $x\in\RR$, $\ket{\psi} = \ket{0}$, then the final state is
\begin{align}
\label{eq:final_state_Rx}
    \ket{\Psi_f} = \ket{\Psi_0} + \ket{\Psi_1}
\end{align}
with 
\begin{align}
    \ket{\Psi_0} &= \sum_{\Vec{j} \in K^n} & p(\Vec{j}) \cos(\frac{1}{2}\sumt{\Vec{x}(\Vec{j})}) \ket{\Vec{j}} \ket{0} \label{eq:psi_0}
    \\
    \ket{\Psi_1} &= \sum_{\Vec{j} \in K^n} & p(\Vec{j}) \sin(\frac{1}{2}\sumt{\Vec{x}(\Vec{j})}) \ket{\Vec{j}} \ket{1} \label{eq:psi_1}
\end{align}
With AE the value $\tilde{a}$ will be estimated, hence $\EE[\cos(S_n)] = 1 - 2\tilde{a}$.
Conversely, if $V(x) = R_y(-x)$, $\ket{\psi} = R_y(\pi/2)\ket{0}$, then there exists a similar decomposition $|\Psi_f'\rangle = \ket{\Psi_0'} + \ket{\Psi_1'}$ so that with $a' = \braket{\Psi'_1}{\Psi'_1}$ we therefore find $\EE[\sin(S_n)] = 1 - 2\tilde{a}'$.
\end{proposition}
The above result opens up an exciting avenue towards a fast Monte-Carlo alternative without the need of sampling strategies. These propositions in conjunction with theorem~\ref{thm:1} which is stated in the following section show that a quantum computer can simulate the quantities $\EE[\cos(S_n)]$ and $\EE[\sin(S_n)]$ efficiently. As a result, one can calculate a random variable's characteristic function $\varphi_{X}(v) = \EE[e^{ivX}] = \EE[\cos(vX)] + i\EE[\sin(vX)]$ with two sets of experiments per $v\in\RR$.

In order to extend the above ideas to estimate expectation values of a range of integrable functions $f: \RR \rightarrow \RR$, a Fourier-series is used. If $f$ is $P$-periodic, then
\begin{equation}
\label{eq:f_fourier_series_approximation}
    f_L(x) = \sum_{l=-L}^L c_l e^{\iu \frac{2\pi l}{P} x}
\end{equation}
is the Fourier-approximation of order $L$ for $f(x)$. By linearity of the expectation value, this approximation carries over to 
\begin{align}
\label{eq:expecation_f_fourier_series_approximation}
    \EE[f_L(S_n)] =\sum_{l=-L}^L c_l \varphi_{S_n}\left(\frac{2\pi l}{P}\right).
\end{align}
As a consequence, it is possible to approximate any such expectation value in $\mathcal{O}(LN)$ experiments, where $N$ is the number of shots per experiment. Convergence on Fourier-series is a rich and mature field~\cite{zygmund2002trigonometric,bary2014treatise,katznelson2004introduction} which establishes basic results about convergence and the rate of convergence of each coefficient for given properties of the function $f$. 

\begin{figure*}[t]
\centering
\subfloat[]{
    \includegraphics[width=0.4\textwidth]{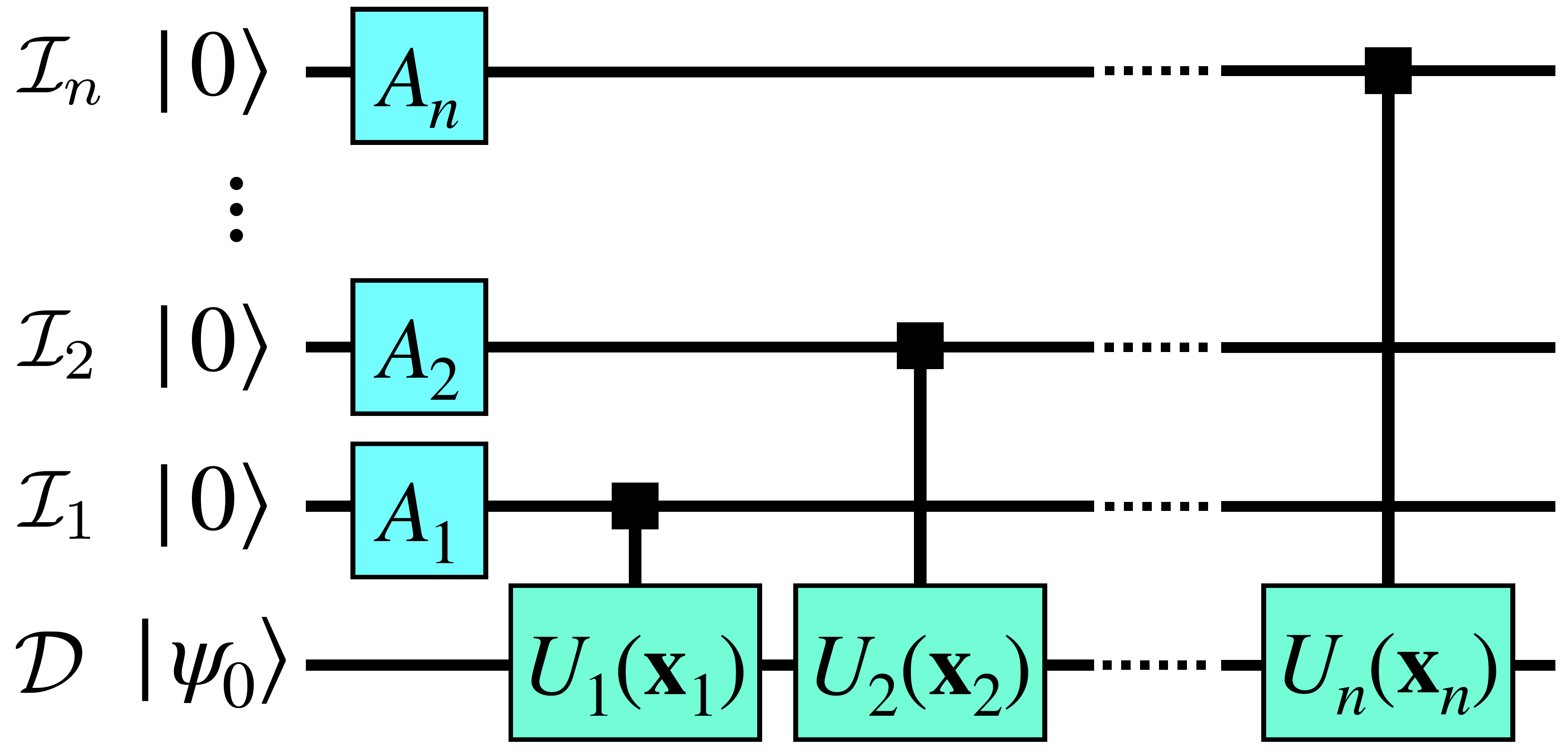}
    }
    \hspace{5mm}
\subfloat[]{
    \includegraphics[width=0.4\textwidth]{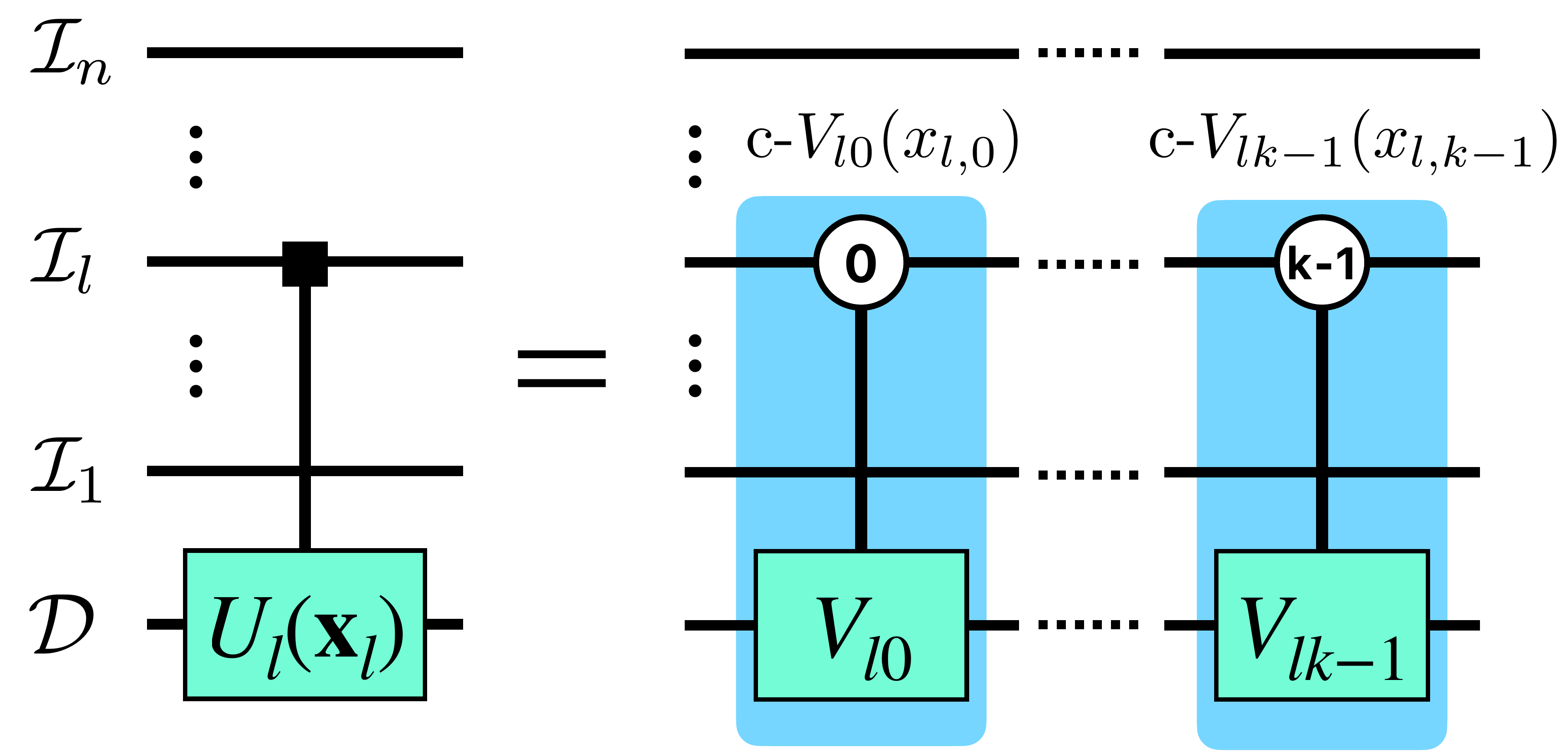}
}
\caption{\textbf{Circuit Design of DSPs with Independent Increments.}
(a) A quantum circuit including preparation of the index system (operators in blue denoted by $A_l$) and the realization of $c\text{-}U(\Vec{j})$ (operators in green denoted by $U_l(\Vec{x})$ with a verticle line connected to black squares). A black square is placed on a control register, and the green box denoted by $U_l(\Vec{x})$is placed on a target register. Unlike the conventional symbol for a controlled-NOT gate between qubits, the black square indicates that one of the $k$ different unitary transformation is performed conditioned on the state of the control register of dimension $k$.
(b) Schematic circuit representation of the applications of $c\text{-}U_l \in B(\mathcal{H}_{\mathcal{I}_l} \otimes \mathcal{H}_\mathcal{D})$ for $l=1, \ldots, n$ as used in (a). We use the notation $V_{lj} = V(x_{l,j})$. The open circle with a number $j$ means that the unitary operator $V_lj$ is applied if the control register state is $|j\rangle$.}
\label{fig:theorem1}
\end{figure*}

The underlying idea of simulating DSPs on a quantum computer has been laid out. Now we show that the simulation can be performed efficiently with a quantum circuit.

%%%===================================
\subsection{Independent increments}
\label{sec:indep_inc}
%%%===================================
To deliver the underlying idea with a simple example, we start by presenting the case for DSPs with independent increments. To this end, the strategy taken is as follows. We first construct the state $\sum_{\Vec{j} \in K^n} p(\Vec{j}) \ket{\Vec{j}} \in \mathcal{H}_\mathcal{I}$, and systematically entangle the data system $\mathcal{D}$ with the index system. When each of the increments are independent of each other, i.e. $\PP[X_i = x, X_j = y] = \PP[X_i = x]\PP[X_j = y]$ for pairwise different $i \not= j$, equation~(\ref{eq:expectation_value_joint_prob}) can be written as
\begin{equation}
\label{eq:expectation_value_joint_prob_independent}
    \EE[f(S_n)] = \sum_{\Vec{j} \in K^n} f(\sumt{\Vec{x}(\Vec{j})}) \prod_{l=1}^n\PP[X_l = x_{l,j_l}],
\end{equation}
and $p^2_{l,j_l} = \PP[X_l = x_{l,j_l}]$.
This structure allows to partition $\mathcal{I}$ into subsystems, so-called \textit{level-index} (sub)systems: $\mathcal{H}_\mathcal{I} = \bigotimes_{l=1}^n \mathcal{H}_{\mathcal{I}_l}$, where each $\mathcal{H}_{\mathcal{I}_l} = \CC^{k}$ represents a qudit Hilbert space. Let $\ket{\alpha(n)} \in \mathcal{H}_\mathcal{I}$ be the index-state. Then it can be described as a product state
\begin{align}
\label{eq:amplitude_to_conditional_probability}
    \ket{\alpha(n)} = \sum_{\Vec{j}} p(\Vec{j}) \ket{\Vec{j}} = \bigotimes_{l=1}^n \left( \sum_{j_l=0}^{k-1} p_{l,j_l}\ket{j_l} \right).
\end{align}
Each evolution from $\ket{\alpha(l)} \rightarrow \ket{\alpha(l+1)}$ is done by an unitary operator $A_{l+1} \in B(\mathcal{H}_{\mathcal{I}_l})$ as given by $A_{l+1} \ket{0} = \sum_{j=0}^{k-1} p_{l+1, j}\ket{j}$. After $n$ levels, i.e. the application of $A= A_n \cdots A_1$, the final state $A \ket{0}_n = \ket{\alpha(n)}$   shown in equation~(\ref{eq:amplitude_to_conditional_probability}) is created. Since each of the operators $A_l$ only operates on a separate subspace $\mathcal{H}_{\mathcal{I}_l}$, they commute and can be applied in parallel. This is a consequence of the independence of the increments $X_l$. For example, an $n$-step DSP with $k=2$ possible paths at each time step can be realized with $n$ index qubits each prepared by a single-qubit unitary operation $A_l\ket{0} = R_y(\theta_l)|0\rangle = \cos(\theta_l/2)|0\rangle + \sin(\theta_l/2)|1\rangle$, where $\theta_l$ is chosen to satisfy $\cos^2(\theta_l/2) = p_{l,0}$ and $\sin^2(\theta_l/2) = p_{l,1}$.

Now, for each step of the DSP, $k$ unitary operators applied to the data system controlled by an index qudit split the data space into $k$ spaces, each attached to an orthogonal subspace of the index system. In other words, in each time step, the data system undergoes $k$ independent trajectories. Thus $n$ steps of $k$ controlled unitary operations allows for the encoding of $k^n$ independent realizations of a DSP to the index-data quantum state. To this end, we identify to each realization $x_{l,j_l}$ of the random variable $X_l$ to the application of the operator $V(x_{l,j_l}) \in B(\mathcal{H}_\mathcal{D})$ to the data system. In fact, the operator will be defined in such a way, that the $l$th index-level state's branches $\ket{j_l}_l$ with the amplitudes $p_{l, j_l}$ (for $j_l \in K$) will be controlling the operator, thereby identifying the probability $p^2_{l, j_l} = \PP[X_l = x_{l, j_l}]$ with the occurrence of $V(x_{l,j_l})$. We denote a projection to the $l$th index subsystem $\mathcal{I}_l$ as 
\begin{equation}
\Pi_{lj} = \ketbra{j}{j}_l \widehat{=} \one_k^{\otimes l-1} \otimes\ketbra{j}\otimes \one_k^{\otimes n-l},
\end{equation}
where $\one_k$ is the $k$-by-$k$ unit matrix and its perpendicular pendant as $\Pi_{lj}^\perp$. Then the following theorem establishes the desired result (see Supplementary Information for the proof).
\begin{theorem}\label{thm:1}
    Let $x\in\RR$ and $j \in K$. Define operators 
    \begin{equation}
    \label{eq:cV}
        \text{c-}V_{lj_l}(x) := \Pi_{lj_l} \otimes V(x) + \Pi_{lj_l}^\perp \otimes I 
    \end{equation}
    with $ \text{c-}V_{lj_l}(x) \in B(\mathcal{H}_{\mathcal{I}_l} \otimes \mathcal{H}_{\mathcal{D}})$. Furthermore, for $l=1, \ldots, n$ and $\mathbf{x} \in \RR^k$, define the operator
    \begin{equation}
    \label{eq:U}
        U_l(\mathbf{x}) = \prod_{j=0}^{k-1} \text{c-}V_{lj_l}(x_j) \in B(\mathcal{H}_\mathcal{I}\otimes\mathcal{H}_\mathcal{D}).
    \end{equation}
    Then $U(\Vec{j}) = U_n(\mathbf{x}_n) \cdots U_1(\mathbf{x}_1)$ for $\Vec{x}_l \in \RR^k,\; l=1, \ldots,n$. The application of $nk$ controlled operations $\text{c-}V_{lj_l}(x_{l,j_l})$ is thus necessary. See Fig.~\ref{fig:theorem1} for a depiction of equations~(\ref{eq:cV}) and~(\ref{eq:U}).
\end{theorem}
Propositions~\ref{prop:1},~\ref{prop:2} and theorem~\ref{thm:1} establish that for $V(x) = R_z(2x)$ or $V(x) = R_y(x)$, $x\in\RR$, we can compute given expectation values of equations~(\ref{eq:expval_sigma_x}) and (\ref{eq:expval_sigma_y}), respectively, with $\mathcal{O}(nk)$ controlled gates. 
\begin{figure*}[t]
\centering
\subfloat[]{
\includegraphics[width=0.45\textwidth]{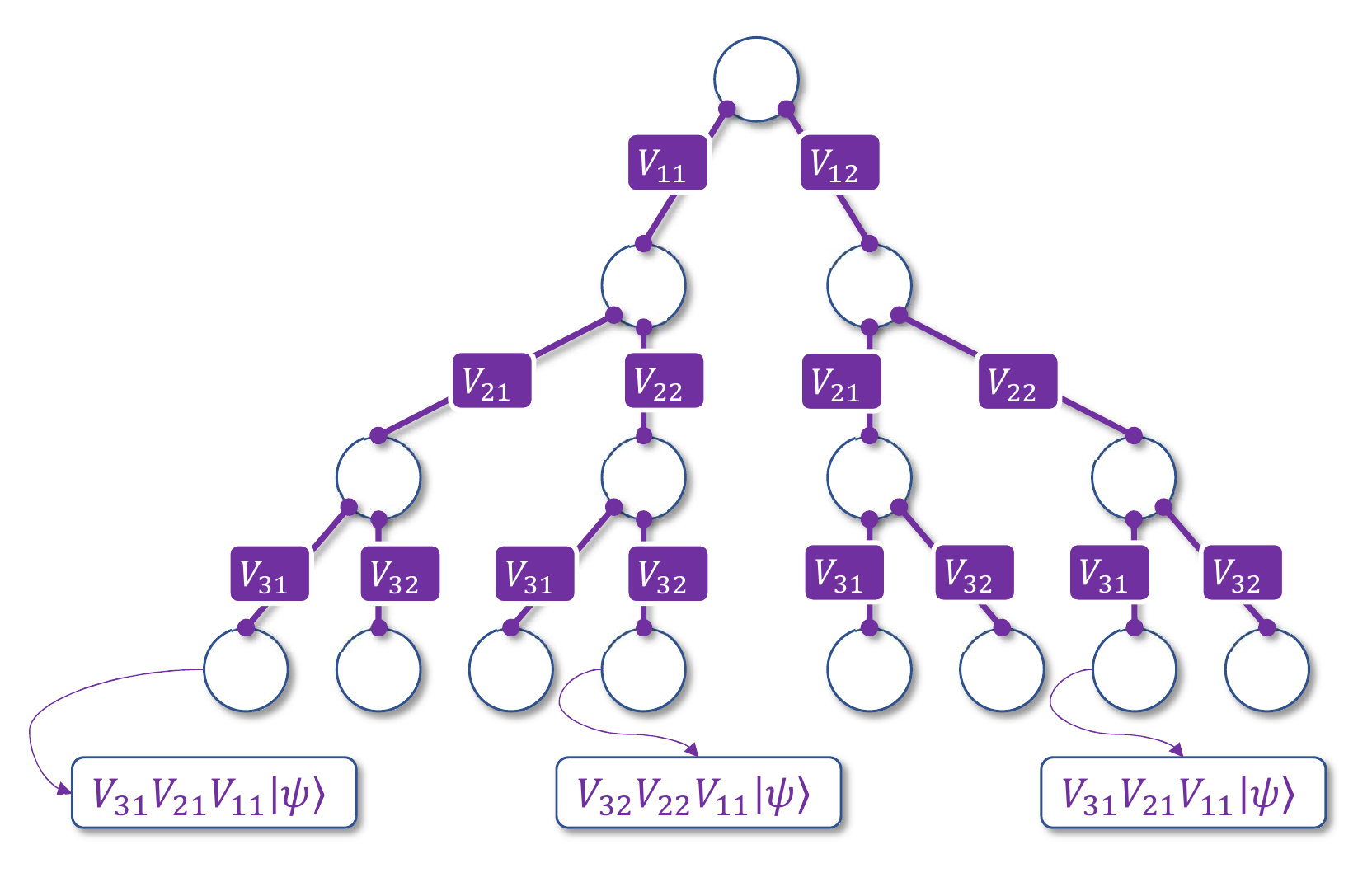}
}
\hspace{5mm}
\subfloat[]{
\includegraphics[width=0.45\textwidth]{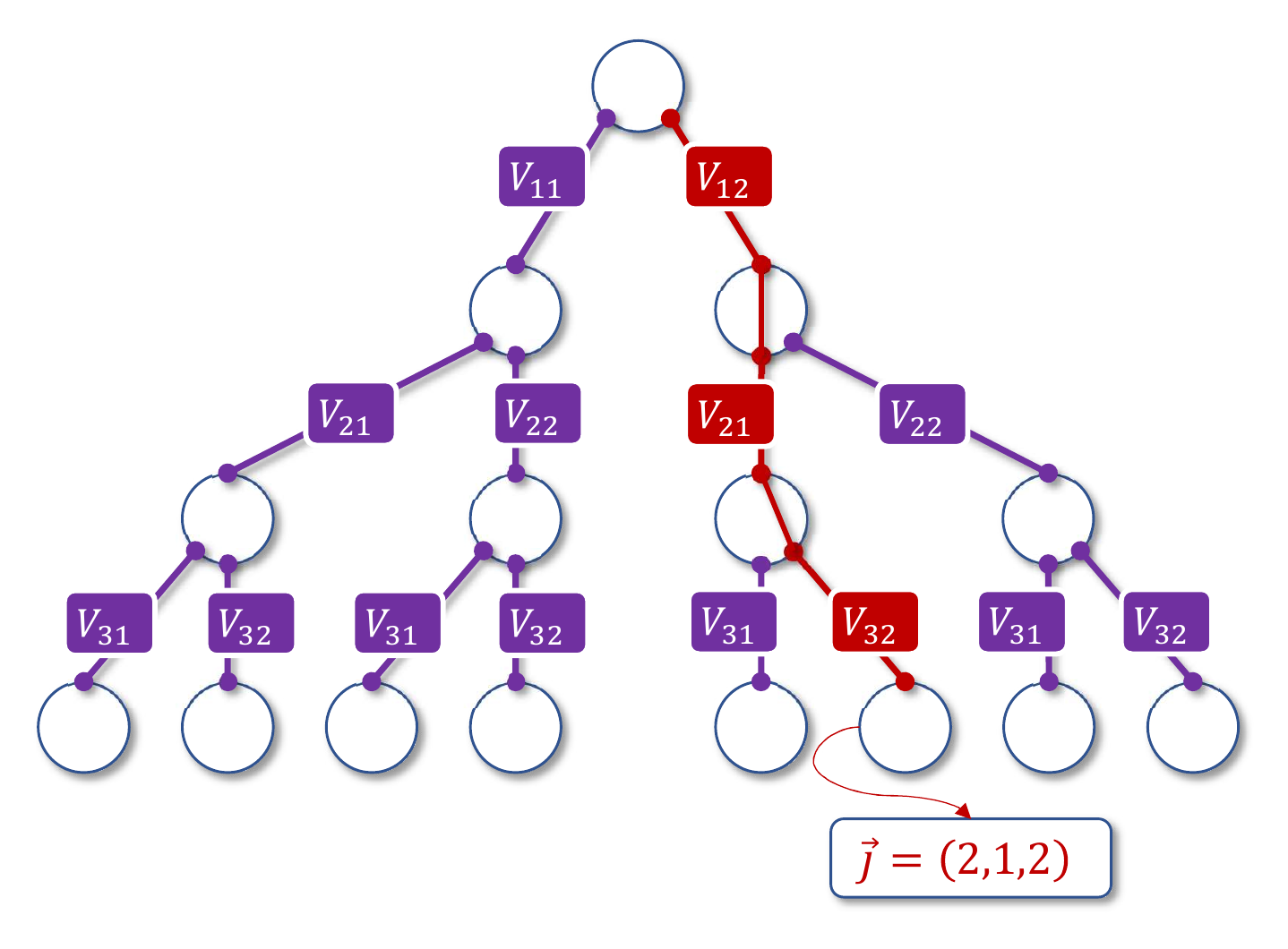}
}
\caption{\textbf{Operator Tree.} (a) The operator tree spanned by a binary example ($k=2$) of $\mathcal{V}$ for three time steps with the edges named. Each edge represents a state transformation from a parent node to a child node, and each node represents a quantum state. Three states at the final leaves starting from an initial state $|\psi\rangle$ at the root node are explicitly shown as an example.  (b) The path $\Vec{j}=(2,1,2)^\top \in K^3$ and the operator $U(\Vec{j}) = V_{32}V_{21}V_{12}$ are shown as an example.}
\label{fig:operator_tree}
\end{figure*}
The above algorithm can be visualized quite intuitively. Consider a set of unitary operators $\mathcal{V} = \{ V_{lj_l} = V(x_{l,j_l}) : l = 1, \ldots, n,\; j_l \in K\}$ as above. These operators span an ordered tree of height $n$ and a maximal degree of $k$. The label of the nodes themselves are of secondary importance, but each edge is labelled by one of the operators of the family $\mathcal{V}$ in such a way that, given the level $l$ ($l = 1$ is the root), each node of the $l$th level has $k$ edges, each of them are labelled by $V_{lj_l},\ j_l\in K$. A path between any two nodes, i.e. $(V_{lj_l}, \ldots, V_{l+ij_{l+i}})$, can be interpreted as a concatenation of operators, i.e. $V(j_l, \ldots, j_{l+i}) = V_{l+ij_{l+i}} \circ \cdots \circ V_{lj_l}$. After $l$ level operations, paths from the root to every nodes at $(l+1)$th level in the operator tree are travelled simultaneously. Hence the data system is evolved by the operator $U(\Vec{j}) = V(j_1, \ldots, j_l) = V_{l j_{l}} \circ \cdots \circ V_{1 j_1}$ for all possible $\Vec{j}$.

%%%===================================
\subsection{Path-dependent increments}
\label{sec:path_dep_inc}
\begin{figure}[t]
    \centering
    \includegraphics[width=0.9\columnwidth]{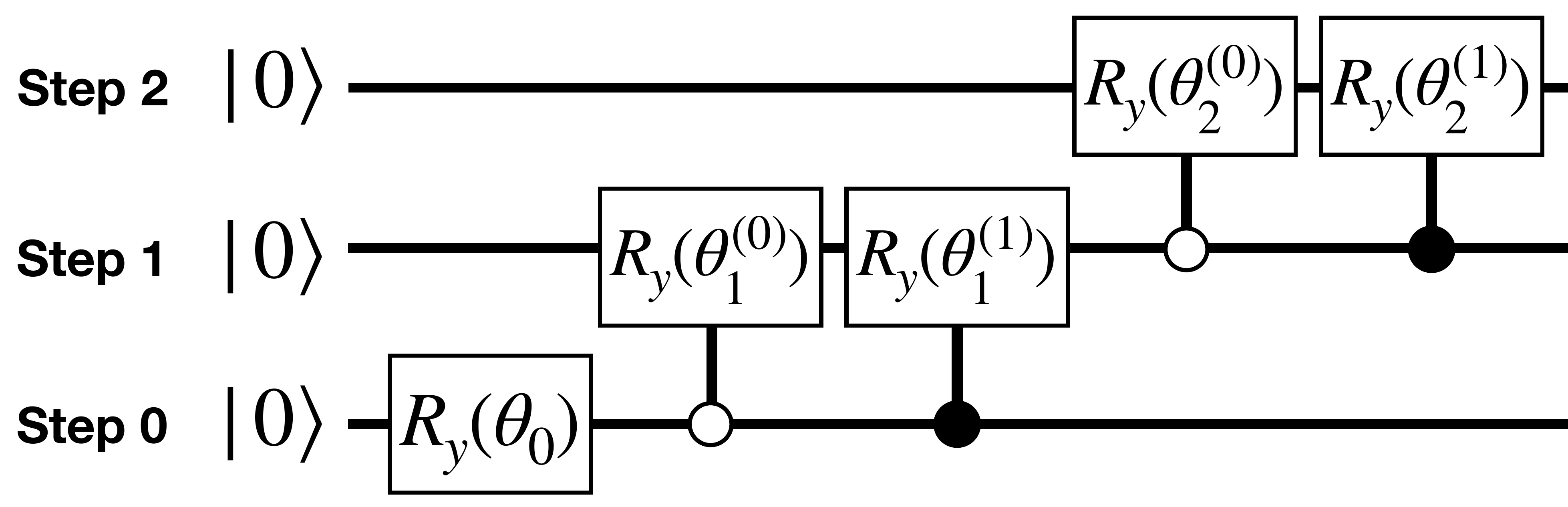}
    \caption{Quantum circuit to prepare the index system for simulating a path-dependent DSP. The figure depicts an example of the first-order Markov chain of two steps each consisting of two realizations.}
    \label{fig:markov_index_circuit}
\end{figure}

%%%===================================
For a DSP with path-dependent increments, the probability of a particular realization can be expressed as
\begin{align}
    \PP[\Vec{X} = \Vec{x}(\Vec{j})] = &\prod_{i=2}^{n}\PP[X_i=x_{i,j_i}|X_{i-1}=x_{i-1,j_{i-1}}]\nonumber\\
    &\times \PP[X_1=x_{1,j_1}].
\end{align}
This is the first-order Markov chain that the probability to take a particular path in a given time step is determined by which path was taken in the previous step. Discussions thus far can easily be extended to such DSPs. The only difference is in the preparation of the index system, i.e. in the construction of the state $\sum_{\Vec{j} \in K^n} p(\Vec{j}) \ket{\Vec{j}}$. Unlike in the case of the independent increments, one needs to introduce entangling operations directly within the index system. Without loss of generality, we use the index system consisting of qubits, i.e. two paths at each level, to describe the procedure. First, the index qubit for the first level is prepared in $A_0|0\rangle=\cos(\theta_0/2)|0\rangle + \sin(\theta_1/2)|1\rangle$ as before, while the rest of the qubits are in $|0\rangle$. Then the second index qubit is prepared by using the controlled operation $|0\rangle\langle 0|\otimes R_y(\theta_2^{(0)})+|1\rangle\langle 1|\otimes R_y(\theta_2^{(1)})$, controlled by the first index qubit, where the superscript indicates the path index of the previous step. Then the second index qubit is used as the control to prepare the third index qubit and so on. Thus, the index system state preparation can be done in $n$ steps and the $l$th index qubit is prepared by applying a controlled operator $|0\rangle\langle 0|\otimes R_y(\theta_l^{(0)})+|1\rangle\langle 1|\otimes R_y(\theta_l^{(1)})$. An example quantum circuit for preparing the index quantum state for a DSP of $n=3$ time steps (levels), each branches to two possibilities, is shown in Fig.~\ref{fig:markov_index_circuit}.

It is also straight-forward to generalize the above procedure to implement path-dependence between events that are more than one time-steps away, by placing the controlled rotation on any two index qubits. Moreover, more complicated path-dependence, such as higher-order Markov chains, can be realized by designing multi-qubit controlled rotations among index qubits.

Besides the index state preparation procedure, the analysis of the DSP follows exactly the same procedure as described in the previous section.

%%%===================================
\subsection{State-dependent increments}
\label{sec:state_dep_inc}
%%%===================================
The increments in a given DSP can also vary in each step, determined by the state in the preceding step. In this case, the probability of a particular realization can be expressed as
\begin{align}
    \PP[\Vec{X} = \Vec{x}(\Vec{j})] = &\prod_{i=2}^{n}\PP[X_i=x_{i,j_i}|\sum_{l=1}^{i-1}X_{l}]\nonumber\\
    &\times \PP[X_1=x_{1,j_1}].
\end{align}
Intuitively, in order to control the increment given a particular realization of the current step, one needs to apply controlled operation to the index register, controlled by the data register in between each successive step. We leave the explicit details on how to realize the above process with quantum circuits as future work.

% ============================================================================
\subsection{Example Applications}
% ============================================================================

% ============================================================================
\subsubsection{The Delta for European call option}
% ============================================================================
An exciting potential application of quantum computing is financial analysis~\cite{rebentrost2018quantumMC,rebentrost2018quantumPortfolio,woerner2019quantum}. In particular, the framework developed in this work can be employed to compute the Delta of an European call option. Let $S_t$ be stochastic process of an underlying asset then
\begin{align}
\label{eq:delta_european_option}
    f(S_t) = \Phi\left( \frac{\ln{\frac{S_t}{K}} + \left(r + \frac{\sigma^2}{2}\right) (T-t) }{\sigma \sqrt{T-t}} \right)
\end{align}
where $\Phi: \RR \rightarrow [0, 1]$ is the cumulative distribution function (CDF) of the standard normal distribution, $K > 0$ is the strike price, $r$ is the risk-free interest rate, $T-t$ is the time to maturity and $\sigma$ is the volatility of the underlying asset. We are interested in the expectation value of the Delta, $\EE[f(S_t)]$ when the underlying asset is described by a geometric Brownian motion, i.e. $S_t = S_0 \exp(\left(\mu - \sigma^2/2\right)t + \sigma W_t)$ where $\mu\in\RR$ and $S_0$ denote the drift and the starting value, respectively. Indeed, equation~(\ref{eq:delta_european_option}) considers a value reminiscent of the log-return, which is modelled by the Brownian motion and can be approximated as a random walk by Donsker's invariance principle~\cite{donsker1951invariance,fristedt2013modern}. The following result establishes the way to implement this on a quantum computer.

\begin{proposition}
    Given $n$ independent and identically distributed random variables $X_l$ with $\PP[X_l = x_1] = \PP[X_l = x_2] = 1/2$ where
    \begin{align}
    x_{1} &=  \frac{\mu - \frac{\sigma^2}{2}}{n\sigma \sqrt{T-t}} - \frac{1}{\sqrt{n} \sqrt{T-t}}
    \\
    x_{2} &=  \frac{\mu - \frac{\sigma^2}{2}}{n\sigma \sqrt{T-t}} + \frac{1}{\sqrt{n}\sqrt{T-t}} 
    \end{align}
    and an initial starting point of 
    \begin{equation}
        x_0 = \frac{\ln S_0 - \ln{K} + \left(r + \frac{\sigma^2}{2}\right) (T-t)}{\sigma \sqrt{T-t}}
    \end{equation}
    with $\tilde{S}_n = x_0 + \sum_{l=1}^n X_l$, we find that
    \begin{align}
        \EE[f(S_t)] = \frac{1}{2} - \sideset{}{'}\sum_{l=-\infty}^{\infty} \frac{\iu}{2\pi l} e^{-2\pi^2 l^2 / P^2} \varphi_{\tilde{S}_n}\left(\frac{2\pi l}{P}\right)
    \end{align}
    when $\Phi$ is limited on the interval $[-P/2, P/2]$. Note that we define the sum with prime as the sum over all summands except for $l=0$ (see Supplementary Information).
\end{proposition}

When using the Pauli measurement scheme as explained in Proposition~\ref{prop:1}, one needs to use $V(vx) = R_z(2vx)$ as operators with constants $v = 2\pi l/P$ for $l=-L, \ldots, L$ to evaluate the characteristic function at those points for the Fourier-series approximation. This makes a total of $2L$ experiments ($L$ measurements for each Pauli observable) with $n$ Hadamard gates on $n$ index qubits for encoding the probability information, $n$ bit-flip gates on the index qubits for implementing the controlled operations that operate if the control qubit is $0$, and $2n$ controlled-$Rz$ gates, which of each can be further decomposed to two controlled-NOT and two $R_z$ gates, applied on a data qubit controlled by the index qubit for implementing $x_{1}$ and $x_{2}$. In addition, the initial start value $x_0$ can be implemented by one $R_z$ gate (see Supplementary Information).

\begin{figure*}[t]
\includegraphics[width=0.85\textwidth]{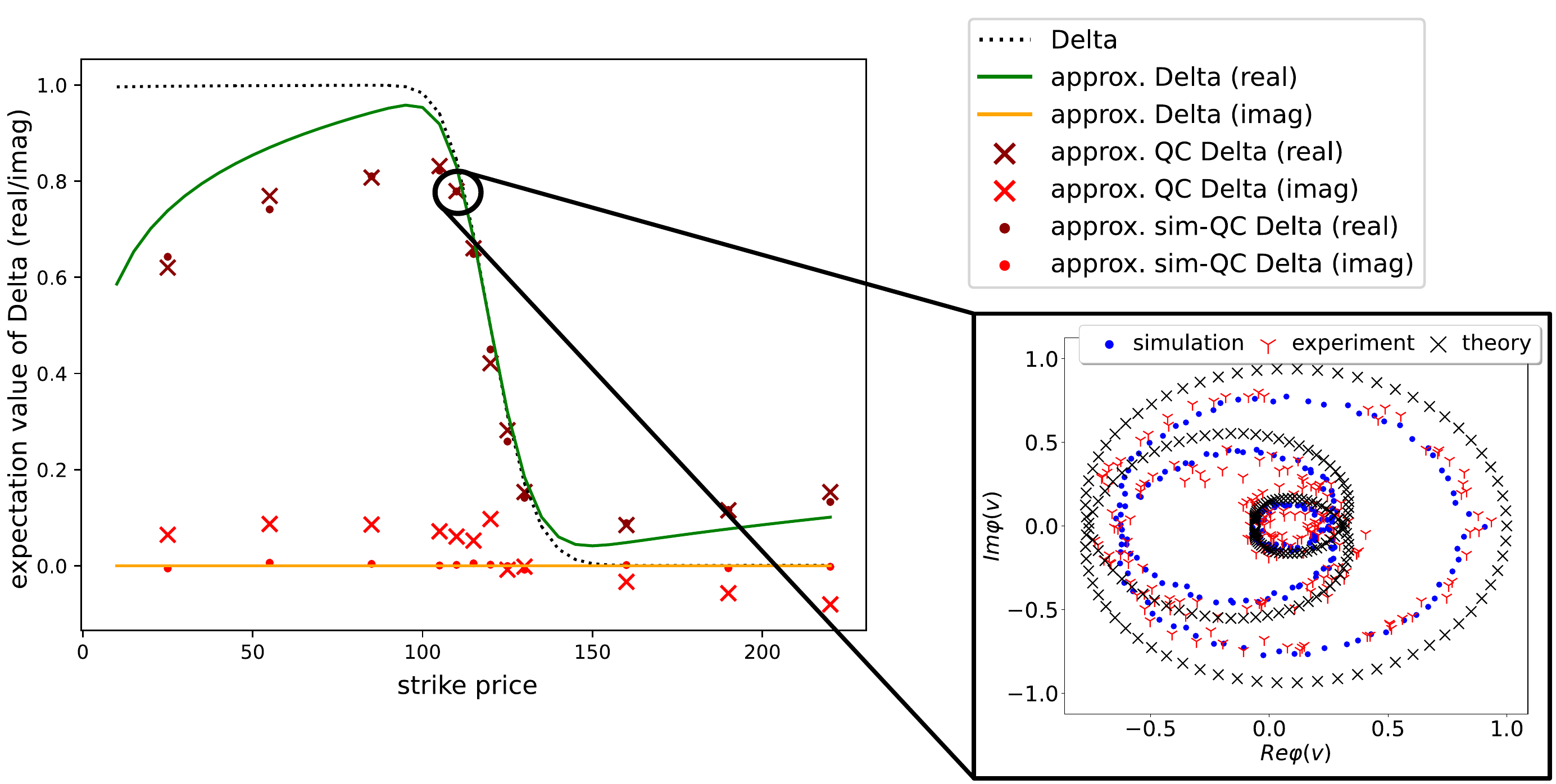}
\caption{\textbf{Evaluation of the Delta for European call option.} (left) The Delta of an European call option is calculated for various strike prices. The underlying asset is defined with $\mu=0, \sigma=0.02, r=0.02$, $S_0=100$, $t=1$ and the time of maturity $T=10$. The dotted black line is the true evaluation of the Delta. The green (orange) solid line is the real (imaginary) part of the theoretical Delta calculation obtained by a Fourier approximation with $P=100$ and $L=100$, which serves as the reference for the experimental validation. The red (brown) crosses are the real (imaginary) part of the Delta calculated with experiment on the IBM quantum computer with error mitigation applied. The dots are the simulation with noise model provided in \texttt{qiskit} with the same error mitigation applied. (right) 
This plot shows the characteristic function calculated in theory ($\times$), by simulation with noise and error mitigation (dot) and the IBM quantum experiment with error mitigation (tri-down) for the example strike price of $K=110$.}
\label{fig:gBm-delta}
\end{figure*}

To demonstrate the proof-of-principle, we performed classical simulations and experiments of the quantum algorithm for calculating the Delta of an European call option on a IBM quantum device named \texttt{ibmqx2}. As an example, the underlying asset is given with $\mu=0$, $\sigma=0.02$, $r=0.02$, $S_0=100$, $t=1$ and the time of maturity $T=10$. The Fourier series approximation is performed by choosing $L=100$ and $P=100$. The standard error mitigation protocol available in \texttt{qiskit}~\cite{Qiskit} is applied to reduce experimental errors. We also performed classical simulation of the quantum algorithm with the standard noise model provided in \texttt{qiskit}. These results are compared with the theoretical values in Fig.~\ref{fig:gBm-delta}. The comparison shows that the noise model provided in \texttt{qiskit} explains the experimental error reasonably well (see Methods for a brief comment on the remaining difference between two results).

Since $\Phi$ is smooth, the Fourier-series coefficients exponentially decays with respect to $l$. However, the choice of the approximation scheme for $\Phi$ comes with the consequence that its approximation $\tilde{\Phi}$ is periodic and continuous at the boundary with $\tilde{\Phi}(P) = \tilde{\Phi}(-P) = 0$. As $P\rightarrow\infty$, the tails extends farther out and flattens to run parallel to the  horizontal axis, thereby converging to $\Phi$. This means that for small $P$ depending on the dynamics of $S_t$ (and hence $\tilde{S}_n$) the approximation may be inaccurate. Therefore, one should keep in mind the minimum/maximum values that $\tilde{S}_n$ can attain and choose $P$ accordingly.

% ============================================================================
\subsubsection{Correlated random walks}
% ============================================================================

As another example, we consider a family of DSPs that strictly requires the ability to simulate path-dependent increments, for which the quantum advantage against classical methods manifests. We demonstrate the simulation of correlated random walks as one such example in the following section.

Correlated random walk (CRW) is a mathematical model that describes discrete random processes with correlations between successive steps~\cite{gillis_1955}. It has been a useful tool to study biological processes~\cite{BOVET1988419,doi:10.1098/rsif.2008.0014}, and can also be used to approximate fractional Brownian motion~\cite{doi:10.1137/1010093,ENRIQUEZ2004203}, which has broad applications for example in mathematical finance~\cite{10.1007/978-3-0348-8291-0_13,arbitrage_fBM,ROSTEK201330} and data network~\cite{400651,10.1007/978-1-4612-4062-4_13,10.1007/978-1-4471-0995-2_14,4673446}. The correlation, often referred to as persistence, results in a local directional bias as the walk moves. More precisely, the CRW denoted by $S_n = \sum_{l=0}^n X_l$ with $n$ discrete random variables $X_l$, $l=1,\ldots,n$ and persistence parameters $p_l\in \lbrack 0,1\rbrack$ and $q_l\in \lbrack 0,1\rbrack$ has the following properties: (1) $X_0=x_0$, (2) $\PP[X_1 = x_1] = \PP[X_1 = x_2] = 1/2$, and (3) $\PP[X_{l+1}=x_1|X_{l}=x_1] = p_l$ and $\PP[X_{l+1}=x_2|X_{l}=x_2] = q_l$ $\forall\; l\ge 1$. The first two properties can be incorporated in quantum simulations easily by following the same procedure used in the previous example. Given $p_l$ and $q_l$, the third property can be implemented as follows. First, all index qubits except the first one that encodes the probability distribution of $X_1$ are initialized in $|0\rangle$. The first index qubit is prepared in $(|0\rangle+|1\rangle)/\sqrt{2}$ in accordance with the second property. Then a controlled rotation gate is applied from each index ancilla qubit to an index qubit of the successive step. The controlled operation can be expressed as
\begin{equation}
    |0\rangle\langle 0 |\otimes R_y(\theta_{p_l}) + |1\rangle\langle 1 |\otimes R_y(\theta_{q_l}),
\end{equation}
where $\theta_{p_l} = 2\cos^{-1}(\sqrt{p_l})$ and $\theta_{q_l} = 2\cos^{-1}(\sqrt{q_l})$. For example, after one application of the above controlled operation, the index qubits representing the probability distribution of the first two steps of the CRW is given as
\begin{equation}
    \frac{|0\rangle\otimes (\sqrt{p_1}|0\rangle + \sqrt{1-p_1}|1\rangle) + |1\rangle\otimes (\sqrt{1-q_1}|0\rangle + \sqrt{q_1}|1\rangle)}{\sqrt{2}}.
\end{equation}
The above state shows that $\PP[X_2=x_1|X_1=x_1]=p_1$ and $\PP[X_2=x_2|X_1=x_2]=q_1$ as required by the property (3) of the CRW.

\begin{figure}[t]
\begin{flushleft}
    \includegraphics[width=0.47\textwidth]{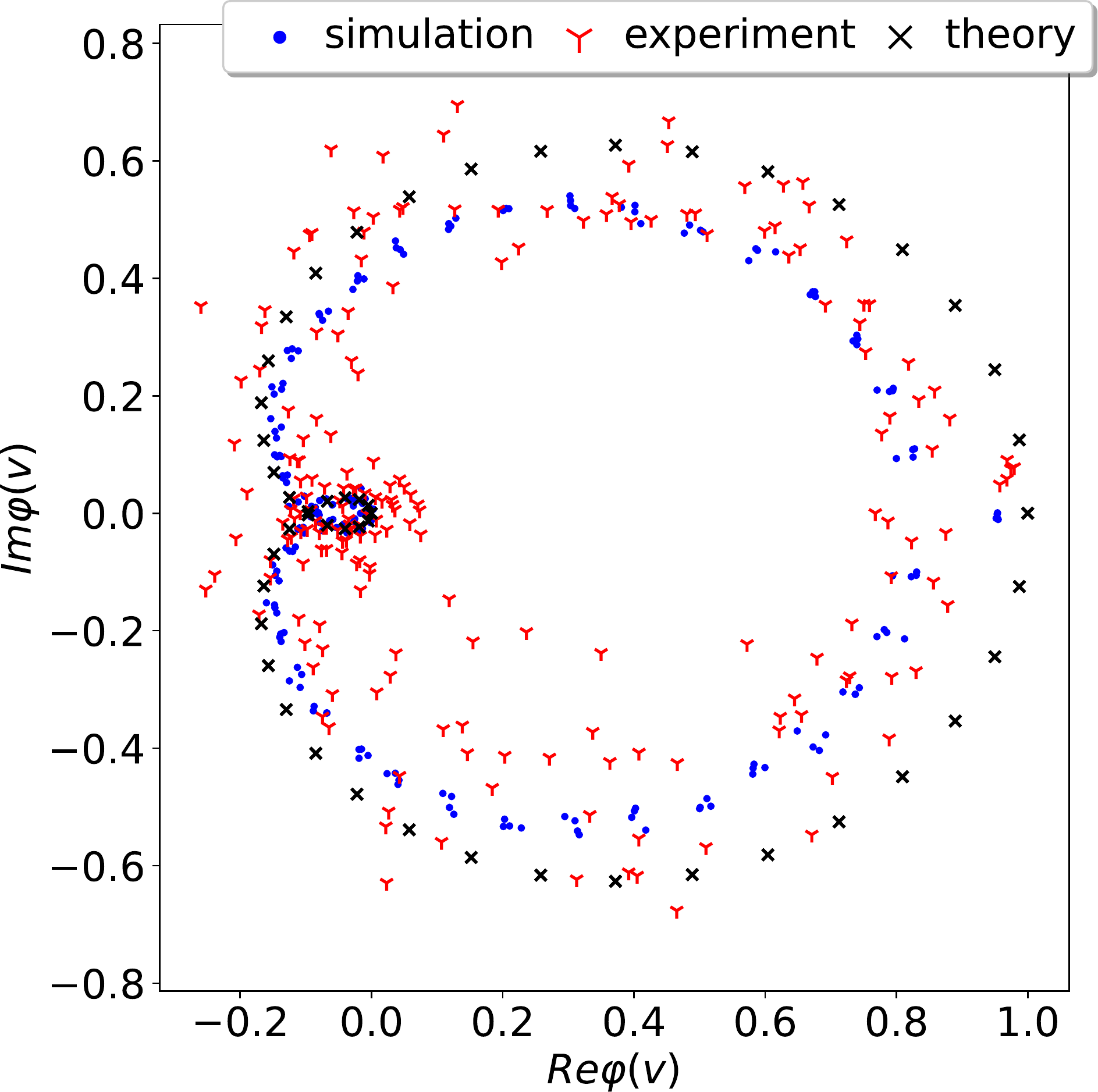}
\end{flushleft}
\caption{\textbf{Characteristic functions of a correlated random walk.} Characteristic functions of a random walk is calculated theoretically ($\times$), by simulation (dot), and by experiment (tri-down). The simulation includes the standard noise model provided in qiskit. The experiment employs the standard error mitigation technique provided in qiskit. The parameters that define the correlated random walk is described in the main text.}
\label{fig:crw}
\end{figure}

We demonstrate the proof-of-principle with an example designed as follows. The correlated random walk is given by increments $x_{l1} = 1$ and $x_{l2} = -1$ with an initial value $x_0 = 0$. The persistence parameters are $p_l = (1/2, 2/3, 5/6, 1)$ and $q_l = (1/2, 1/3, 1/6, 0)$ with $l=1,2,3,4$. Experiments were performed to calculate the characteristic function for $v_l = 2 \pi l/P$, $l=-L, \ldots, L$ with $L=100$ and $P=100$ on \texttt{ibmqx2}. The standard error mitigation protocol available in \texttt{qiskit} is applied to reduce experimental errors. We also performed classical simulation of the quantum algorithm with the standard noise model provided by \texttt{qiskit}. These results are compared with the theoretical values in  Fig.~\ref{fig:crw}.

% ====================================================================
\section{Discussion}
% ====================================================================
We presented a quantum-classical hybrid framework for estimating an expectation value of any integrable function of a random variable in discrete stochastic process with independent increments. As the main ingredient of the framework, we developed a quantum algorithm for efficiently calculating point evaluations of the characteristic function of a random variable, which may also lead to other interesting applications since the probability distribution of a random variable can be completely defined by its characteristic function. More specifically, in the quantum part, the framework proposes a succinct representation of a classical DSP with independent increments as a quantum state in the form of equation~(\ref{eq:process_state}). The joint probability and the value of each realization are encoded in an entangled state $\ket{\Psi_f}$, and therefore all necessary information about a DSP is present. We also detailed the construction of a quantum circuit for preparing the quantum state $\ket{\Psi_f}$ using the number of circuit elements that only grows linearly with the total number of time steps. For a DSP of $n$ total steps each consisting of $k$ possibilities, there are $k^n$ paths. Such process can be encoded in a quantum state using $n$ $k$-dimensional index qudits (or $\lceil\log_2(k)\rceil$ qubits) and $nk$ controlled gates. There is no need for sampling strategies as all realizations exist in quantum superposition. This fact is exploited in the steps afterwards. 

Two different measurement schemes are proposed for the estimation of expectation values $\EE[\cos(S_n)]$ and $\EE[\sin(S_n)]$. The first scheme is to measure an expectation value of $\sigma_x$ and $\sigma_y$ directly on the data system of $\ket{\Psi_f}$, resulting in a convergence error of $\mathcal{O}(1/\sqrt{N})$ for $N$ repeated experiments for any $N>0$. This shows that the quantum brute-force approach acquires the optimal importance distribution. Furthermore, the sampling convergence can be improved by utilizing the quantum amplitude estimation technique, which promises to reduce the approximation error by a factor of $\mathcal{O}(1/2^m)$ using $m$ additional qubits. However, the resource overhead of the latter approach is $m$ ancilla qubits, $\mathcal{O}(\mathrm{poly}(m))$ Grover-like controlled operations, and $\mathcal{O}(m^2)$ one- and two-qubit gates for implementing QFT. Therefore, with the noisy intermediate-scale quantum (NISQ) devices, it may be desirable to use the former approach.

The advantage of this algorithm lies in two facts. First, it is a quantum-classical hybrid computation and thus is a viable candidate to be solved with near-term quantum devices. One can envisage multiple near-term devices running in parallel to calculate all $2L$ terms independently at the same time. Similar parallelization is also suitable for multiple partitions of a large quantum devices where the qubit connectivity is high within each partition but low among partitions. Second, given one type of process $S_n$, we can pre-compute a number of evaluations of the characteristic functions $\varphi_{S_n}(\pm v_i)$ for $v_1 < \cdots < v_L \in \RR$ \textit{a priori}. With those evaluations at hand, it is possible to assemble the Fourier-series on-demand when given the Fourier-coefficients of a function $f$. This approach makes it possible to invest resources to maximize the precision of the characteristic function evaluation for regularly used DSPs. Furthermore, our framework promotes the idea for a co-design (special-purpose) quantum computer~\cite{brown2016co,langford2017experimentally,lamata2018digital,parra2020digital}, which focuses on optimizations and design decisions at the hardware level to particularly support the DSP simulations at hand.

We underscore our findings with two interesting examples. First, we showed an application to finance, the calculation of the Delta of a European call option. The key idea was to model the stochastic behavior of an underlying asset as a Brownian motion, which is then approximated as a random walk by Donsker's invariance principle. Next, an application to DSPs with path-dependent increments is demonstrated by an example of correlated random walks. We performed and presented the results of proof-of-principle experiments for each example to demonstrate the validity and the feasibility of our method.

The framework can be extended to multi-variate functions $f: \RR^d \rightarrow \RR$. Given $P_1, \ldots, P_d \in\RR$, the period of the respective argument of $f$ and $L_1, \ldots, L_d \in \NN_+$, a multi-dimensional Fourier-series approximation
\begin{equation*}
f(x_1, \ldots, x_d) = \sum_{l_1 = 1}^{L_1} \cdots \sum_{l_d=1}^{L_d} c(\omega_1, \ldots, \omega_d) e^{\iu \sum_i \frac{2\pi l_i}{P_i} x_i}
\end{equation*}
can be applied like equation~(\ref{eq:f_fourier_series_approximation}) and consequently equation~(\ref{eq:expecation_f_fourier_series_approximation}). The primary harmonics to calculate are thus $\varphi_{S_n}(\Vec{v}) = \EE[e^{\iu \Vec{v}\cdot \Vec{S_n}}]$,
% \begin{align*}
%     \varphi_{S_n}(\Vec{v}) &= \EE[e^{\iu \Vec{v}\cdot \Vec{S_n}}] 
%     \\
%     &= \EE[\cos(\Vec{v}\cdot \Vec{S_n})] + \iu \EE[\sin(\Vec{v}\cdot \Vec{S_n})],
% \end{align*}
which can be achieved by increasing the dimension of the index systems and applying a new set of a unitary family $\mathcal{V}_r$ ($r=1, \ldots, d$) to the same data system $\mathcal{D}$.

The ability to simulate multi-variate stochastic dynamics with quadratic quantum speed-up without requiring any sampling strategies is highly beneficial for the study of discrete stochastic processes. The path-dependent and state-dependent DSP simulations are excellent fits for studying random walks with internal states~\cite{hughes1995random}, and discrete processes that converge to Ornstein-Uhlenbeck~\cite{PhysRev.36.823} and fractional Brownian motion~\cite{doi:10.1137/1010093}. Due to the broad applicability of these mathematical models, the framework developed in this work presents tremendous opportunities for solving problems that arise from various disciplines, such as physics, biology, epidemiology, hydrology, engineering, and finance.

In future work, we plan to provide explicit quantum circuit design for simulating state-dependent DSPs. Interesting applications to accompany the future work include the analysis of stochastic epidemic models~\cite{TUCKWELL200776,ALLEN2017128} and of hot streak~\cite{hot_streak_nature2018}.

% ====================================================================
\section*{Methods}
% ====================================================================

All experiments are performed using one of the publicly available IBM quantum devices consisting of five superconducting qubits, named as \texttt{ibmqx2}. In order to fully utilize the IBM quantum cloud platform, we used the IBM quantum information science kit (\texttt{qiskit}) framework~\cite{Qiskit}. The versions---as defined by PyPi version numbers---used for this work were 0.20.0.

Superconducting quantum computing devices that are currently available via the cloud service, such as those used in this work, have limited coupling between qubits. Resolving coupling constraints as well as optimizations are done in \texttt{qiskit} with a preset of so-called \textit{pass managers}. The optimization level ranges from 0 to 3. As we chose the \texttt{ibmqx2} for our experiments the mapping of data register and index register follow the connectivity. For the family of devices that contains \texttt{ibmq\_ourense}, one swap operation must be used to exchange the physical qubit 3 and 4. For both layouts see Supp. Info. Figure 3a and 3b. During the compilation step of the experiments we first use a level 0 pass and then a subsequent level 3 pass to optimize and resolve connectivity constraints. To reduce experimental errors, we used the standard error mitigation functionality of \texttt{qiskit}. This requires an extra set of experiments to be executed prior to the main experiment.

In order to understand the source of experimental discrepancy, the experimental results were compared to simulation results that take a realistic noise model into account. During the execution of an experiment, the current device parameters were gathered and stored. Upon completion, a simulation was executed with the standard \texttt{qiskit} noise model, also applying error mitigation on the result. The remaining discrepancy between experimental and simulation results can be attributed to errors that are not included in the \textit{basic error model}, such as various cross-talk effects, drift, and non-Markovian noise. The standard noise model is described in the supplementary information of Ref.~\cite{blank2020quantum} in detail.

The example for the Delta applied the strike price $K$ from 10 to 220 in increments of 5 for the theoretical calculation, while the experiments were performed for $K=25, 55, 85, 105, 110, 115, 120, 125, 130, 160, 190, 220$. The experiment for each $K$ is executed for characteristic function evaluations at $v_l = 2\pi l / P$ with $l = -L, \ldots, L$, $L=100$ and $P=100$, each with 8129 shots. For the correlated random walk experiment, we used 32786 shots for each of the evaluations of $v_l = 2\pi l / P$ with $l = -L, \ldots, L$ and $P=L=100$. Note that this example did not use the Fourier approximation, and $P$ has been chose to be the same as $L$ so that $v_l$ goes through one period on each side, resulting in a symmetric picture. As the IBM quantum cloud platform allow for 8192 shots per execution, we created multiple identical experiments and manually added the results.

% ===========================================================================================
% ===========================================================================================
\subsection*{Data availability}
% ===========================================================================================
% ===========================================================================================
% The datasets generated during and/or analysed during the current study are available on the GitHub repository~\cite{SupplementaryRepository}.
The data that support the findings of this study are available from C.B. upon reasonable request.

\section*{Acknowledgements}
We acknowledge use of IBM Q for this work. The views expressed are those of the authors and do not reflect the official policy or position of IBM or the IBM Q team. 
We thank Philipp Leser for fruitful discussions on financial applications.
This research is supported by the National Research Foundation of Korea (Grant No. 2019R1I1A1A01050161 and 2018K1A3A1A09078001), by the Ministry of Science and ICT, Korea, under an ITRC Program, IITP-2019-2018-0-01402, and by the South African Research Chair Initiative of the Department of Science (UID: 64812) and Technology and the National Research Foundation.
\linebreak
\section*{Author contributions statement}
C.B. and D.K.P. contributed equally to this work. C.B. and D.K.P designed and analysed the model. 
All authors reviewed and discussed the analyses and results, and contributed towards writing the manuscript. F.P. is the corresponding author.\\
\linebreak
\textbf{Competing interests} The authors declare no competing interests.

% \bibliographystyle{unsrt}
% \bibliography{references}
%

%
\end{document}

\title{Supplementary Information: Quantum-enhanced analysis of discrete stochastic processes}

\author{Carsten Blank}
\email{blank@data-cybernetics.com }
\affiliation{Data Cybernetics, 86899 Landsberg, Germany}
\author{Daniel K. Park}
\email{dkp.quantum@gmail.com}
\affiliation{School of Electrical Engineering, KAIST, Daejeon, 34141, Republic of Korea}
\affiliation{ITRC of Quantum Computing for AI, KAIST, Daejeon, 34141, Republic of Korea}
\author{Francesco Petruccione}
\email{petruccione@ukzn.ac.za}
\affiliation{School of Electrical Engineering, KAIST, Daejeon, 34141, Republic of Korea}
\affiliation{Quantum Research Group, School of Chemistry and Physics, University of KwaZulu-Natal, Durban, KwaZulu-Natal, 4001, South Africa}
\affiliation{National Institute for Theoretical Physics (NITheP), KwaZulu-Natal, 4001, South Africa}

\begin{abstract}

\end{abstract}

\maketitle
\def\one{{\mathchoice {\rm 1\mskip-4mu l} {\rm 1\mskip-4mu l} {\rm \mskip-4.5mu l} {\rm 1\mskip-5mu l}}}

% ====================================================================
\section{Expectation Value of Equation~(1)}
% ====================================================================

The expectation value $\EE[f(S_n)] = \sum_s f(s) \PP[S_n = s]$ describes the value of a realization $s$ of $S_n$ and its probability. We exploit the structure of an incremental process such as a discrete stochastic process (DSP). By elementary calculations we find
\begin{align*}
&\EE[f(S_n)] = \sum_s f(s) \PP[S_n = s] \nonumber 
\\    
&= \sum_s \left(\sum_{x_{1,j_l} + \cdots + x_{n,j_n} = s} f\left(\sum_{l=1}^n x_{l,j_l}\right) \PP\left[\bigcap_{l=1}^n X_l = x_{l,j_l} \right]\right) \nonumber 
\\
&= \sum_{\substack{x_{l,j_l} \in B_l;\\j_l=0,\ldots,k-1,\\ l = 1, \ldots, n}} f\left(\sum_{l=1}^n x_{l,j_l}\right) \PP\left[\bigcap_{l=1}^n X_l = x_{l,j_l} \right] 
\\
&= \sum_{\Vec{j} \in K^n} f(\sumt{\Vec{x}(\Vec{j})}) \PP[\Vec{X} = \Vec{x}(\Vec{j})]
\end{align*}
which shows the validity of equation~(1) of the main text.

% ====================================================================
\section{Index Register Calculation}
% ====================================================================

By separating the probability amplitude of $p(\Vec{j}) = \prod_{l=1}^n p_{l, j_l}$ we show that the product state is achieved:
\begin{align}
    \ket{\alpha(n)} &= \sum_{\Vec{j}} p(\Vec{j}) \ket{\Vec{j}} \nonumber
    \\
    &= \sum_{j_1, \ldots, j_n} p_{1,j_1} \cdots p_{n, j_n} \ket{j_1} \otimes \cdots \otimes \ket{j_n} \nonumber 
    \\
    &= \sum_{j_1, \ldots, j_n} p_{1, j_1} \ket{j_1} \otimes \cdots \otimes  p_{n, j_n} \ket{j_n} 
    \\
    &= \bigotimes_{l=1}^n \left( \sum_{j=1}^k p_{l,j_l}\ket{j} \right).
\end{align}

% ====================================================================
\section{Proof of Propositions~1 and~2}
% ====================================================================

The proposition 1 states that the expectation value measurements of Pauli X and Y, i.e. $\sigma_x$ and $\sigma_y$, observables coincide with the evaluation of the expectation values of $\EE[\cos(S_n)]$ and $\EE[\sin(S_n)]$ if the unitary acting on the data system is a phase gate.
\begin{proof}
Let $V(x) = R_z(2x)$ (which is linear in the parameter). We then find
\begin{equation}
    \label{eq:unitary_u1}
    V(x_{i,j_i}) = R_z(2x_{i,j_i}) = \ketbra{0}{0} + e^{i x_{i,j_i}} \ketbra{1}{1}
\end{equation}
by factoring out a global phase. Then for the unitary operation $U(\Vec{j}) = R_z(2\sum_{i=1}^n x_{i,j_i})$ and the observable $M=\sigma_x$, we find
\begin{align}
    M(\Vec{j}) &= U^\dagger(\Vec{j}) M U(\Vec{j}) \nonumber
    \\
    &= e^{-i \sum_{i=1}^n x_{i,j_i}} \ketbra{1}{0} + e^{i \sum_{i=1}^n x_{i,j_i}} \ketbra{0}{1}.
\end{align}
So if we start with an equal superposition state, i.e. $\ket{\psi} = (\ket{0} + \ket{1})/\sqrt{2}$, we end up 
with $\expval{M(\Vec{j})} = \cos(\sum_{i=1}^n x_{i,j_i})$. Hence equation~(4) of the main text becomes
\begin{equation}
\label{eq:sigma_x_expectation_value}
    \expval{\sigma_x} = \sum_{\Vec{j} \in B} p^2(\Vec{j}) \cos(\sum_{i=1}^n x_{i,j_i})
\end{equation}
Likewise, measuring the expectation value of $M=\sigma_y$ yields 
\begin{equation}
    M(\Vec{j}) =\quad e^{-\iu \pi/2 - \iu \sum_{i=1}^n x_{i,j_i}} \ketbra{1}{0} + i e^{\iu \pi/2 + i \sum_{i=1}^n x_{i,j_i}} \ketbra{0}{1},
\end{equation}
and again if the initial state is an equal superposition state, we find $\expval{M(\Vec{j})} = \sin(\sum_{i=1}^n x_{i,j_i})$. 
\end{proof}

In proposition 2, instead of using the phase gate, we use the single-qubit rotation gate around the y-axis of the Bloch sphere as basis of the unitary family $\mathcal{V}$. Then the final state of equation~(3) in the main text is separated to two orthogonal subspaces with cosines or sines as amplitudes. The result is pretty simple and derives from direct computation as in the proof of Proposition 1, but with $U(x) = R_y(x)$.
\begin{proof}
Let $V(x_{i,j_i}) = R_y(x_{i,j_i}) = \cos(\frac{1}{2}x_{i,j_i}) I - \iu \sin(\frac{1}{2}x_{i,j_i}) \sigma_y$  -- which is linear in the parameter -- we then find $U(\Vec{j}) = R_y(\sum_{i=1}^n x_{i,j_i})$ and therefore
\begin{align*}
    U(\Vec{j}) = \cos(\frac{1}{2}\sum_{i=1}^n x_{i,j_i}) I - \iu \sin(\frac{1}{2}\sum_{i=1}^n x_{i,j_i}) \sigma_y.
\end{align*}
So if we start with the ground state ($\ket{\psi} = \ket{0}$) we find according to equation~(3) of the main text
\begin{align}
    \ket{\Psi_f} &= \sum_{\Vec{j} \in K^n} p(\Vec{j}) \ket{\Vec{j}} \otimes \left(\cos(\frac{1}{2}\sum_{i=1}^n x_{i,j_i}) \ket{0} + \sin(\frac{1}{2}\sum_{i=1}^n x_{i,j_i}) \ket{1}\right) \nonumber
    \\
    &= \underbrace{\sum_{\Vec{j} \in K^n} p(\Vec{j}) \cos(\frac{1}{2}\sum_{i=1}^n x_{i,j_i}) \ket{\Vec{j}} \ket{0}}_{\ket{\Psi_0}}
    + \underbrace{\sum_{\Vec{j} \in K^n} p(\Vec{j}) \sin(\frac{1}{2}\sum_{i=1}^n x_{i,j_i}) \ket{\Vec{j}} \ket{1}}_{\ket{\Psi_1}}.
    \label{eq:AE_state}
\end{align}
Additionally, we see by the same argument that starting with the excited state will reverse the roles of cosine and sine.

The measurement scheme of AE now will estimate a value $a=\braket{\Psi_1}{\Psi_1} = \sum_{\Vec{j} \in K^n} p^2(\Vec{j}) \sin^2(\frac{1}{2}\sum_{i=1}^n x_{i,j_i})$, by trigonometric identities. This is equivalent to
\begin{align*}
    a &=\braket{\Psi_1}{\Psi_1} = \sum_{\Vec{j} \in K^n} p^2(\Vec{j}) \frac{1}{2}\left(1 - \cos(\sum_{i=1}^n x_{i,j_i}) \right)
    \\
    &= \frac{1}{2} \sum_{\Vec{j} \in K^n} p^2(\Vec{j})  - \frac{1}{2} \sum_{\Vec{j} \in K^n} p^2(\Vec{j}) \cos(\sum_{i=1}^n x_{i,j_i}) 
    \\
    &= \frac{1}{2}  - \frac{1}{2} \sum_{\Vec{j} \in K^n} p^2(\Vec{j}) \cos(\sum_{i=1}^n x_{i,j_i})
    \\
    &= \frac{1}{2}  - \frac{1}{2} \EE[\cos(S_n)].
\end{align*}
Thus $\EE[\cos(S_n)] = 1 - 2a$, which was to be shown. The other result is given when setting $V(x_{i,j_i}) = R_y(-x_{i,j_i})$ and having the initial state $\ket{\psi} = R_y(\pi/2)$. This results in a operator $U(\Vec{j}) = R_y(\pi/2 - \sum_{i=1}^n x_{i,j_i})$, and hence negates the angle and adds a pre-factor. Thus
$$
a' = \frac{1}{2}  - \frac{1}{2} \sum_{\Vec{j} \in K^n} p^2(\Vec{j}) \cos(\frac{\pi}{2} - \sum_{i=1}^n x_{i,j_i})
$$
As a result,
$$
a' = \frac{1}{2}  - \frac{1}{2} \EE[\cos(\frac{\pi}{2} - S_n)] = \frac{1}{2}  - \frac{1}{2} \EE[\sin(S_n)].
$$
This concludes the proof: $\EE[\sin(S_n)] = 1 - 2a'$.
\end{proof}

% ====================================================================
\section{Amplitude Estimation}
% ====================================================================

The technique of amplitude estimation (AE) uses phase estimation to find the amplitude of a state. It's use was originated by quantum search algorithms~\cite{brassard2002quantum} in which the technique of Grover's search is extended to amplitude amplification to the ``good" states. In short, say $\mathcal{A}$ is an algorithm/operator creating a final state $\mathcal{A}\ket{0} = \ket{\Psi_f} = \ket{\Psi_0} + \ket{\Psi_1}$ which is separated in two orthogonal subspaces. The operator applied is similar to the revert around the mean of Grover's search
$$
Q = Q(\mathcal{A}, f) = - \mathcal{A} S_0 \mathcal{A}^\dagger S_f
$$
where $f: \ZZ \rightarrow \{0,1\}$, $S_0 = -\ketbra{0}{0} + \sum_{\Vec{j} \not= 0} \ketbra{\Vec{j}}{\Vec{j}}$ and
$$
S_f \ket{\Vec{j}} = \begin{cases}
-\ket{x}, f(x) = 1
\\
\quad  \ket{x}, f(x) = 0
\end{cases}.
$$
This means that $S_0 = S_f$ where $f(0) = 1$ and $f(x) = 0$ for all $x \not= 0$. Every such binary function $f$ partitions the Hilbert space $\mathcal{H} = \mathcal{H}_1 \oplus \mathcal{H}_2$ in two subspaces, which lead to the canonical partitioning of $\ket{\Psi_f}$ into $\ket{\Psi_0}$ and $\ket{\Psi_1}$. In the situation of the main text, we have equation~(\ref{eq:AE_state}): 
$$
    \ket{\Psi_0} = \sum_{\Vec{j} \in K^n} p(\Vec{j}) \cos(\frac{1}{2}\sum_{i=1}^n x_{i,j_i}) \ket{\Vec{j}} \ket{0} 
$$
and
$$
    \ket{\Psi_1} = \sum_{\Vec{j} \in K^n} p(\Vec{j}) \sin(\frac{1}{2}\sum_{i=1}^n x_{i,j_i}) \ket{\Vec{j}} \ket{1}.
$$
Therefore, the binary function is very simple; if we interpret $x = j_1 \cdots j_n j_{n+1}$ as a binary number with the least significant bit being $j_{n+1}$, then we find the correct function to be
$$
    f(x) = \begin{cases}
    1,\; x \text{ even}
    \\
    0,\; x \text{ odd}
    \end{cases}.
$$
The value to be under scrutiny is $a = \braket{\Psi_1}{\Psi_1}$ and in particular $\sin^2{\theta_a} = a$. It is known that the eigenvalues of $Q$ are $\lambda_\pm = e^{\pm \iu 2 \theta_a}$. Then we define the operator $\Lambda_M(U): \ket{j}\ket{y} \mapsto \ket{j}(U^j \ket{y})$ for $0 \leq j < M$. Given an $M > 0$, usually a power of two, the algorithm uses $m = \log_2 M$ qubits: 
$$
    (F_m^\dagger \otimes I_{\mathcal{H}}) \Lambda_M(Q) (H^{\otimes m} \otimes I_{\mathcal{H}}) \ket{0} \otimes \ket{\Psi_f}
$$
where $F_m$ is the quantum Fourier transform. The implementation of $\Lambda_M(Q)$ is usually done by controlled operators
$$
\prod_{j=0}^{m-1} \texttt{c-}Q^{2^j}_{j1} = \Pi_{j1} \otimes Q^{2^j} + \Pi^\perp_{j1} \otimes I_{\mathcal{H}}
$$
with $\Pi_{jl}$ as given in equation~(16) of the main text. Measuring an observable $\sigma_z^{\otimes m} \otimes I_{\mathcal{H}}$ yields one outcome $y$ with probability $\frac{8}{\pi^2}$, namely the estimate $\tilde{a} = \sin{(\pi \frac{y}{M})}$ of $a$, and the error is given by $|a - \tilde{a}| < \mathcal{O}(M^{-1})$~\cite{brassard2002quantum}.

% ====================================================================
\section{Proof of Theorem 1}
% ====================================================================

Theorem 1 establishes the efficient construction of the family of operators $U(\Vec{j})$ for $\Vec{j} \in \RR^k$. Given an initialized index system $\mathcal{I}$ we need to show
\begin{equation}
\label{eq:full_evolution_operators}
    U_n(\mathbf{x}_n) \cdots U_1(\mathbf{x}_1) \ket{\alpha(n)} \otimes \ket{\psi} = \ket{\Psi_f} = \sum_{\Vec{j} \in K^n} p(\Vec{j}) \ket{\Vec{j}} \otimes U(\Vec{j}) \ket{\psi}
\end{equation}
according to equation~(3) of the main text for a DSP.

\begin{proof} 
We will establish the proof by a simple inductive argument. To see that equation~(\ref{eq:full_evolution_operators}) coincides with equation~(3) of the main text, we look into the first level evolution:
\begin{align*}
     U_1(\mathbf{x}_1) \ket{\alpha(1)} \otimes \ket{\psi}
     &= U_1(\mathbf{x}_1) \sum_{j=1}^k \left(p_{1j}\ket{j}_1\right) \otimes \ket{\psi} 
     \\
     &= \prod_{j=1}^k \text{c-}V_{1j}(x_{1j}) \sum_{j=1}^k \left(p_{1j}\ket{j}_1\right) \otimes \ket{\psi} 
     \\
     &= \sum_{j=1}^k p_{1j}\ket{j}_1 \otimes V(x_{1j}) \ket{\psi}.
\end{align*}
In summary, $k$ orthogonal subspaces have been popularized by the application of $A_1$ and on each such subspace the data systems undergoes an evolution of $V(x_{1j})$.

Now let assume we were able to create the $l$th level, then we show that this also holds true
\begin{align*}
     &U_{l+1}(\mathbf{x}_{l+1}) U_l(\mathbf{x}_l) \cdots U_1(\mathbf{x}_1) \ket{\alpha(l+1)} \otimes \ket{\psi} 
     \\
     &= U_{l+1}(\mathbf{x}_{l+1}) \sum_{\Vec{j} \in K^l} p(\Vec{j}) \ket{\Vec{j}} \otimes A_{l+1}\ket{0} \otimes U(\Vec{j}) \ket{\psi}
     \\
     &= U_{l+1}(\mathbf{x}_{l+1}) \sum_{\Vec{j} \in K^l} \sum_{j=0}^{k-1} p(\Vec{j}) p_{l+1,j} \ket{\Vec{j}} \otimes \ket{j} \otimes U(\Vec{j}) \ket{\psi}
     \\
     &= \prod_{j=1}^k \text{c-}V_{l+1,j}(x_{l+1,j}) \sum_{\Vec{j} \in K^l} \sum_{j=0}^{k-1} p(\Vec{j}) p_{l+1,j} \ket{\Vec{j}} \otimes \ket{j} \otimes V(x_{l,j_l}) \cdots V(x_{1,j_1}) \ket{\psi}
     \\
     &= \sum_{\Vec{j} \in K^l} \sum_{j=0}^{k-1} p(\Vec{j}) p_{l+1,j} \ket{\Vec{j}} \otimes \ket{j} \otimes V(x_{l+1,j}) V(x_{l,j_l}) \cdots V(x_{1,j_1}) \ket{\psi}
     \\
     &= \sum_{\Vec{j} \in K^{l+1}} p(\Vec{j}) \ket{\Vec{j}} \otimes V(x_{l+1,j_l}) V(x_{l,j_l}) \cdots V(x_{1,j_1}) \ket{\psi}
     \\
     &= \sum_{\Vec{j} \in K^{l+1}} p(\Vec{j}) \ket{\Vec{j}} \otimes U(\Vec{j}) \ket{\psi}
\end{align*}
At each step there are exactly $k$ applications of the operator $\text{c-}V_{lj}$ and there are $n$ levels to compute. Thus we need $nk$ operators. The number of branches is $k^n$ however. This exponential reduction only works as we do the evolution of a set of operators $\text{c-}V_{lj}$ on all subspaces simultaneously without knowing or distinguishing the current branch we evolve from. Certainly, this is a major restriction to what can be achieved if this procedure is merely taken as a state preparation routine. However, it resembles a DSP and therefore the specific structure can be easily exploited. 

After this remark, we can formally conclude the proof.
\end{proof}

% ====================================================================
\section{Simulation of the Brownian Motion by Donsker's Invariance Principle}
% ====================================================================

The Brownian Motion is defined as 
\begin{equation}
    B_t = \mu_B t + \sigma_B W_t
\end{equation}
where $\mu_B \in \RR$ denotes the drift, $\sigma_B \in \RR_+$ denotes the volatility and $W_t$ denotes the Wiener process. By Donsker's invariance principle~\cite{donsker1951invariance,fristedt2013modern} a Wiener process is approximated by a random walk. In our situation let $X_l$ be an independent and identically distributed (i.i.d.) increments ($l=1, \ldots, n$) with $\PP[X_l = -1] = \PP[X_l = +1] = \frac{1}{2}$ and define
$$
X_{\frac{m}{n}}^{(n)} = \frac{1}{\sqrt{n}} \sum_{l=1}^m X_l
$$
and define $X^{(n)}$ to all of $[0,1]$ by being linear on each sub-interval $I_m = (\frac{m-1}{n}, \frac{m}{n}]$, i.e.
$$
X_t^{(n)} = \sum_{m=1}^n X_{\frac{m}{n}}^{(n)} \ONE_{I_m}(t).
$$
The distribution of $X^{(n)}$ converges to the distribution of the Wiener process $W$ on $[0,1]$ (Donsker's Invariance Principle). From this we define
\begin{align}
    B_t^{(n)} &= \mu_B t + \sigma_B X_t^{(n)} \nonumber
    \\
    &=  \mu_B t + \sigma_B \sum_{m=1}^n X_{\frac{m}{n}}^{(n)} \ONE_{I_m}(t) \nonumber
    \\
    &=  \sum_{m=1}^n \left( \frac{\mu_B}{n} t + \sigma_B X_{\frac{m}{n}}^{(n)} \ONE_{I_m}(t) \right).
\end{align}
In our case, the value $t>0$ is a fixed quantity. Thus the last equation reduces to
\begin{align}
     B_t^{(n)} &= \sum_{l=1}^m \left( \frac{\mu_B}{n}t + \sigma_B \frac{1}{\sqrt{n}} X_l \right) \\
     &= \frac{m}{n} \mu_B t +  \sigma_B \frac{1}{\sqrt{n}} \sum_{l=1}^m X_l
\end{align}
with $m$ such that $t \in I_m$. As such, we define $Y_l = \frac{\mu_B}{n}t + \sigma_B \frac{1}{\sqrt{n}} X_l$ with
\begin{align}
\label{eq:simulation_brownian_motion_random_walk}
 \PP[Y_l = \frac{\mu_B}{n}t - \sigma_B \frac{1}{\sqrt{n}}] = \PP[Y_l = \frac{\mu_B}{n}t + \sigma_B \frac{1}{\sqrt{n}}] = \frac{1}{2}.
\end{align}

% ====================================================================
\section{Proof of Proposition~3}
% ====================================================================

In order to simulate a geometric Brownian motion in the context of a Delta of a European option, we notice that we are interested in a value similar to the log-return. For this reason, the underlying asset can be described by the Brownian motion. As described above, with the help of Donsker's Invariance Principle, it is possible to approximate the Brownian motion with a random walk. Proposition 4 of the main manuscript states how the random walk must be chosen. Then the Fourier-coefficients of the cumulative distribution function (CDF) of the standard normal distribution must be approximated in order to compute the expectation value of the Delta of an European option.
\begin{proof}
The process to approximate is given by equation~(21) of the main text
\begin{align*}
    f(S_t) = \Phi\left( \frac{\ln{\frac{S_t}{K}} + \left(r + \frac{\sigma^2}{2}\right) (T-t) }{\sigma \sqrt{T-t}} \right)
\end{align*}
the goal is to only Fourier approximate $\Phi(\cdot)$ and redefining the process that is being simulated. We assume a geometric Brownian motion of the underlying asset given by
$$
    S_t = S_0 \exp(\left(\mu - \frac{\sigma^2}{2}\right)t + \sigma W_t)
$$
and since we are interested in the term $\ln S_t$, we find
$$
    \ln S_t = \ln S_0 + \left(\mu - \frac{\sigma^2}{2}\right)t + \sigma W_t.
$$
By redefining the process according to 
\begin{align}
\tilde{S}_t &= \frac{\ln S_t}{\sigma \sqrt{T-t}} + \frac{-\ln{K} + \left(r + \frac{\sigma^2}{2}\right) (T-t)}{\sigma \sqrt{T-t}}
\\
&= \frac{1}{\sigma \sqrt{T-t}} \left( \left(\mu - \frac{\sigma^2}{2}\right)t + \sigma W_t \right) + \frac{\ln S_0 - \ln{K} + \left(r + \frac{\sigma^2}{2}\right) (T-t)}{\sigma \sqrt{T-t}}
\\
&= \frac{\mu - \frac{\sigma^2}{2}}{\sigma \sqrt{T-t}}t + \frac{1}{\sqrt{T-t}} W_t + \frac{\ln S_0 - \ln{K} + \left(r + \frac{\sigma^2}{2}\right) (T-t)}{\sigma \sqrt{T-t}}
\end{align}
with the drift $\mu_B = \frac{\mu - \frac{\sigma^2}{2}}{\sigma \sqrt{T-t}}$, volatility $\sigma_B=\frac{1}{\sqrt{T-t}}$ and starting point 
$$
    x_0 = \frac{\ln S_0 - \ln{K} + \left(r + \frac{\sigma^2}{2}\right) (T-t)}{\sigma \sqrt{T-t}}.
$$ 
This defines a Brownian motion. By applying equation~(\ref{eq:simulation_brownian_motion_random_walk}) with aforementioned drift and volatility, we find
\begin{align}
    x_{l1} &=  \frac{\mu - \frac{\sigma^2}{2}}{n\sigma \sqrt{T-t}} - \frac{1}{\sqrt{n} \sqrt{T-t}}
    \\
    x_{l2} &=  \frac{\mu - \frac{\sigma^2}{2}}{n\sigma \sqrt{T-t}} + \frac{1}{\sqrt{n}\sqrt{T-t}} 
\end{align}
for $l=1, \ldots, n$. Since all paths are equally probable, the index system must be initialized to an equal superposition state by applying $H^{\otimes n}$ to $\ket{0}^{\otimes n}$. The first operation on the data system is a rotation by the angle $x_0$. After that, the operators $U_l(\Vec{x}_l)$ with $\Vec{x}_l = (x_{l1}, x_{l2})^\top$ are each applied for $l=1, \ldots, n$, confer equation~(18) of the main manuscript.

Last to show are the Fourier-coefficients of the CDF of the standard normal distribution. In fact, given the error function
\begin{equation}
    \erf(x) = \int_0^x e^{-t^2} dt,
\end{equation}
the CDF of the standard normal distribution is then $\Phi(x) = \frac{1}{2}\left((1 + \erf(x/\sqrt{2})\right)$. Using the Fourier transform $\mathcal{F}$, we find that $\mathcal{F}(\erf)(t) = -\iu e^{-\pi^2 t^2} / (\pi t)$ and thus, in the sense of distributions, 
\begin{align*}
    \mathcal{F}(\Phi)(t) &= \frac{1}{2} \mathcal{F}(1)(t) + \frac{1}{2} \mathcal{F}\left(\erf(\frac{\cdot}{\sqrt{2}})\right)(t) 
    \\
    &=\frac{1}{2} \delta_0(t) + \frac{1}{2} \frac{-\iu e^{-2 \pi^2 t^2}}{\pi t}
    \\
    &= \hat{f}(t).
\end{align*}
By applying the limiting process from the Fourier transform to Fourier series, we find that given $\hat{f}$ we find that the Fourier coefficients
$$
    c_n = \frac{1}{P} \int_{-\frac{T}{2}}^{\frac{T}{2}} f(x)e^{-2\pi \iu \left(\frac{n}{P}\right)x} dx
$$
correspond to the rule 
$$
    c_n = \frac{1}{P} \hat{f}\left(\frac{n}{P}\right)
$$
under the assumption that $\Phi$ is zero outside of the interval $[-T/2, T/2]$ without any discontinueties. As $T\rightarrow \infty$ this will approach the correct limit. We therefore find
\begin{align}
    c_n &= -\iu \frac{1}{2\pi n} e^{-2\pi^2 n^2 / P^2}, \quad n \not= 0
    \\
    c_0 &= \frac{1}{2}
\end{align}
which was to be shown.
\end{proof}

% ====================================================================
\section{Alternative Solution to Expectation value of a CDF}
% ====================================================================

Suppose we want to estimate a quantity $\EE[f(S)]$ for some random variable $S$, where $f(S) = \Phi(y(S))$ and $\Phi(\cdot)$ is the CDF of a standard normal distribution. Thus $\Phi(y(S)) = \left(1+\frac{2}{\sqrt{\pi}}\int^{y(S)/\sqrt{2}}_0 \exp(-t^2) dt\right)/2$. Then
$$
\EE[f(S)] = \frac{1}{2} + \frac{1}{\sqrt{\pi}}\EE[\int^{y(S)/\sqrt{2}}_0 \exp(-t^2) dt].
$$
Now, from earlier results, we know that $\exp(-t^2)$ can be fourier-approximated as $\exp(-t^2)\approx \sum a_l \cos(l\pi t/P)$ with $a_l$ calculated in the previous section. Thus, the expectation value that we need to calculate becomes
\begin{align}
\EE[\int^{y(S)/\sqrt{2}}_0 \exp(-t^2) dt] & \approx \EE[\int^{y(S)/\sqrt{2}}_0 \frac{\sqrt{\pi}}{P} + \sum_{l=1}^L a_l \cos(l\pi t/P) dt]\nonumber \\
& = \EE[\frac{\sqrt{\pi}y(S)}{\sqrt{2}P} + \sum_{l=1}^L a_l \int_0^{y(S)/\sqrt{2}}\cos(l\pi t/p)dt] \nonumber \\
& = \frac{\sqrt{\pi}}{\sqrt{2}P}\EE[y(S)] +\sum_{l=1}^L a_l \EE[\int_0^{y(S)/\sqrt{2}}\cos(l\pi t/p) dt]\nonumber \\
& = \frac{\sqrt{\pi}}{\sqrt{2}P}\EE[y(S)] + \sum_{l=1}^L a_l \EE[\frac{P\sin(l\pi y(S)/\sqrt{2}P)}{l\pi}].
\end{align}
The expectation values in the second term of the above equation can be estimated by the standard procedure introduced in our work. On the other hand, the first term can be further approximated using the Fourier series, and the problem of finding the Fourier-coefficients depends on the function $y$.

% ====================================================================
\section{Simulation of the Delta}
% ====================================================================

\begin{figure}[ht]
\includegraphics[width=0.8\textwidth]{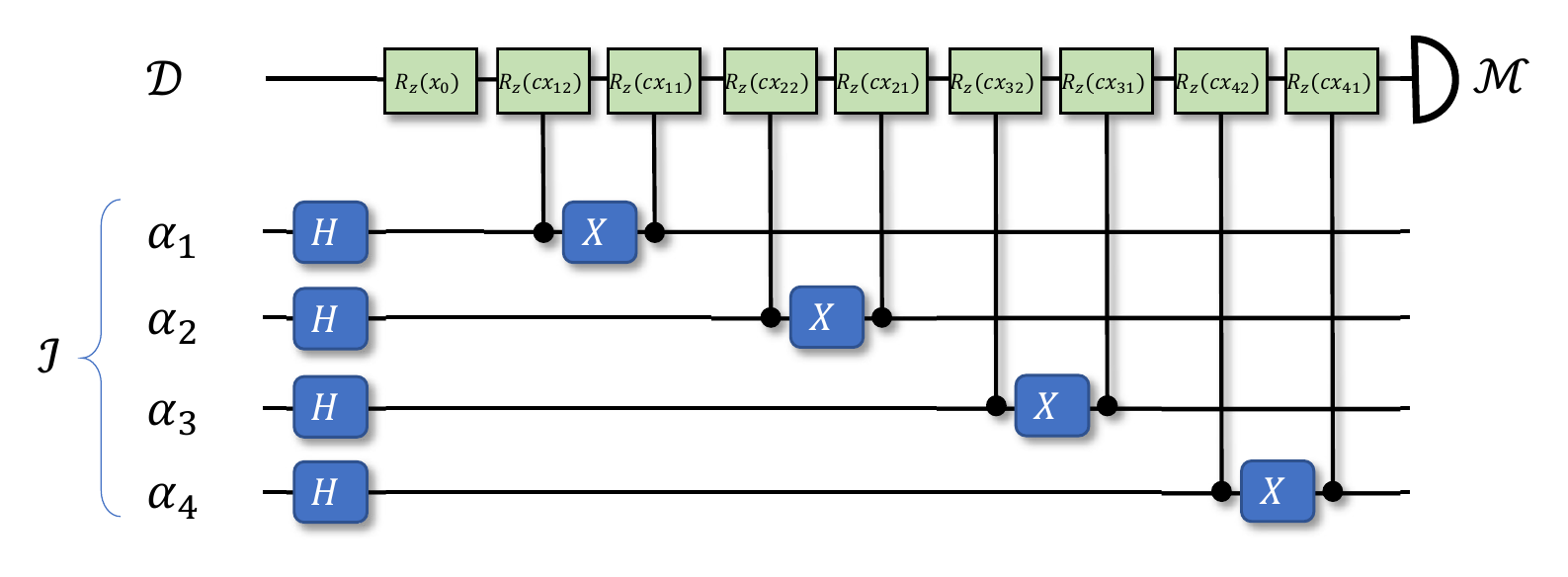}
\caption{\textbf{Quantum circuit for estimating the Delta of European option.} We have four levels each with its own set of angles to rotate. Different values of $c$ will be used to compute the characteristic function $\varphi_{S_4}(v)$ when two set of experiments are done, one with an observable $\mathcal{M} = \sigma_x$ and the other one with an observable $\mathcal{M} = \sigma_y$.}
\label{fig:numerical_integration_circuit}
\end{figure}

\begin{figure}[ht]
\subfloat[]{
\includegraphics[width=0.25\textwidth]{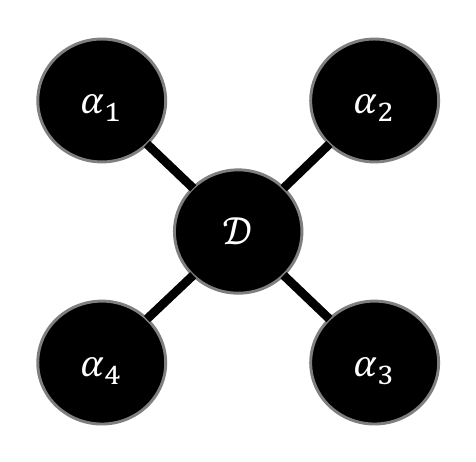}
}
\hspace{2cm}
\subfloat[]{
\includegraphics[width=0.25\textwidth]{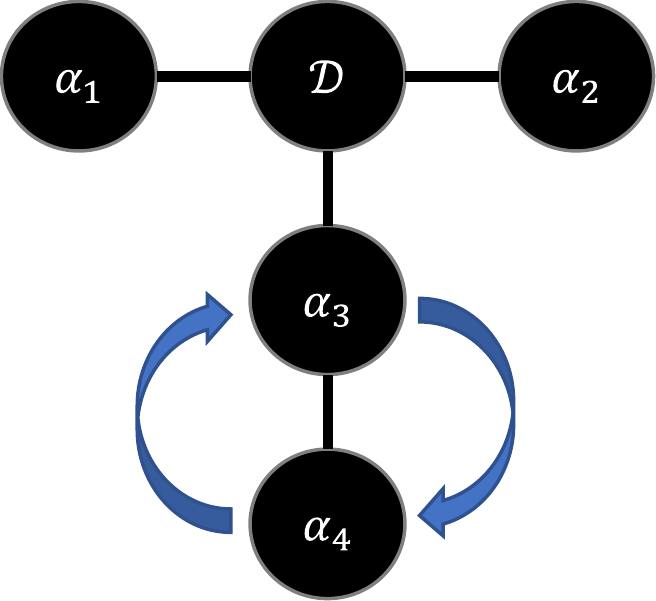}
}
\caption{\textbf{Devices' Layout.} Shown are the coupling maps of the two five-qubit IBM quantum device we used in our experiments. Each node represents a qubit and each edge represents the connectivity. Every index subsystems needs to be connected to the data system, the \texttt{ibmqx2} (left) is a choice that fulfills the necessary connections are naturally given by the connectivity while the \texttt{ibmq\_ourense} (right) will need one swap operation (indicated by the blue arrows) for $\alpha_4$ to be able to operate on the data register $\mathcal{D}$.}
\label{fig:ibmqx_connectivity}
\end{figure}

We want to simulate
\begin{align*}
    \EE[f(S_t)] = \EE[\Phi\left( \frac{\ln{\frac{S_t}{K}} + \left(r + \frac{\sigma^2}{2}\right) (T-t) }{\sigma \sqrt{T-t}} \right)]
\end{align*}
with $S_t$ being a geometric Brownian motion as given in the main text. The simulation on a real quantum device, such as \texttt{ibmqx2} (Suppl. Figure~\ref{fig:ibmqx_connectivity}) puts the restriction on the maximum number of time steps one can simulate due to the number of qubits that are available. With this five-qubit device for example, one can study the random walk
$$
\tilde{S}_4 = x_0 + X_1 + X_2 + X_3 + X_4
$$
with $X_l$ i.i.d. given as in proposition 4 of the main text. The circuit to calculate the characteristic function $\varphi_{S_4}(v)$ is given by Suppl. Figure~\ref{fig:numerical_integration_circuit} and is therefore equal to the circuit to the numerical calculation example of proposition 3 of the main manuscript except the choices of $x_0$ and $x_{lj}$ ($l=1,2,3,4,\ j=1,2$). We calculate the characteristic function at $L=5$ positions, so $v_1 = \frac{\pi}{10}, v_2=\frac{\pi}{5}, v_3 = \frac{3\pi}{10}, v_4 = \frac{2\pi}{5}, v_5 = \frac{\pi}{2}$. For each position, two expectation value measurements are needed, i.e. $\mathcal{M} = \sigma_x$ and with $\mathcal{M} = \sigma_y$. We find that by setting together the coefficients and the characteristic function evaluations
$$
\frac{1}{2} - \sum_{l=1}^5 \left( \iu \frac{1}{2\pi n} e^{-2\pi^2 n^2 / P^2} \varphi_{\tilde{S}_4}(n) + c.c.\right) \approx \EE[f(S_t)].
$$

\begin{figure}[ht]
\includegraphics[width=0.4\textwidth]{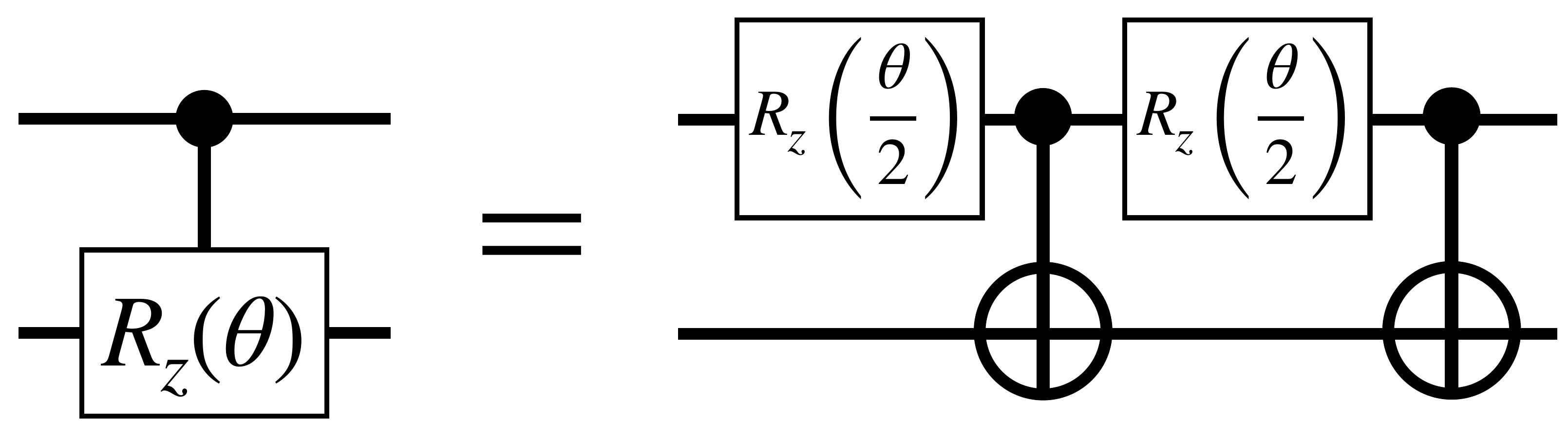}
\caption{\textbf{Decomposition of the controlled $R_z$ gate.} A controlled single-qubit rotation around the $z$ axis of the Bloch sphere, i.e. $R_z$, gate can be decomposed to two single-qubit $R_z$ gates and two controlled-NOT gates as depicted in the figure.}
\label{fig:controlled_rz_gate}
\end{figure}

The corresponding quantum circuit is shown in Figure~\ref{fig:numerical_integration_circuit} and shows two important facts. The first is that the index registers are all initialized by a Hadamard gate to put each qubit in an equal superposition state. This corresponds to the probability of occurrence of each outcome. The second is, that controlled $R_z$ gates are easily implemented on a real quantum device, for example, a five-qubit quantum device \texttt{ibmqx2} provided by IBM via the cloud, using two phase gates (these are usually virtual) and only two \texttt{cnot} gates each (refer to Figure~\ref{fig:controlled_rz_gate}). Looking at the connectivity of this particular quantum processor unit (QPU) as an example, refer to Figure~\ref{fig:ibmqx_connectivity}, this also means that \texttt{swap} operations are not necessary. In total, there are 16 controlled-NOT gates, 4 single-qubit bit-flip (X) gates, 4 Hadamard gates and 16 $Rz$ gates in use.

\bibliographystyle{unsrt}
% \bibliography{suppl_references}